\newif\ifeapj 
\eapjtrue 

\ifeapj 
  \documentclass[iop]{emulateapj}
  \submitted{}
\else 
  \documentclass[manuscript]{aastex}
  \newenvironment{deluxetable*}[1]{\begin{deluxetable}{#1}}{\end{deluxetable}}
\fi

\usepackage{amsmath}
\usepackage{placeins}
\newcommand{\vect}[1]{\textbf{\textit{#1}}}
\newcommand{\pdm}[2]{\frac{\partial #1}{\partial #2}}
\newcommand{\sn}[2]{#1 \times 10^{#2}}

\renewcommand{\ion}[2]{#1~\textsc{\romannumeral#2}}

\shorttitle{Model for Solar and Stellar Flares}
\shortauthors{Allred, Kowalski, and Carlsson}

\begin{document}

\title{A Unified Computational Model for Solar and Stellar Flares}
\author{Joel C. Allred}
\affil{NASA/Goddard Space Flight Center, Code 671, Greenbelt, MD 20771}
\email{joel.c.allred@nasa.gov}
\author{Adam F. Kowalski}
\affil{Department of Astronomy, University of Maryland College Park}
\and
\author{Mats Carlsson}
\affil{Institute of Theoretical Astrophysics, University of Oslo, P.O. Box 1029, Blindern N-0315, Oslo, Norway}

\begin{abstract}
We present a unified computational framework which can be used to describe impulsive flares on the Sun and on dMe stars. The models assume that the flare impulsive phase is caused by a beam of charged particles that is accelerated in the corona and propagates downward depositing energy and momentum along the way. This rapidly heats the lower stellar atmosphere causing it to explosively expand and dramatically brighten. Our models consist of flux tubes that extend from the sub-photosphere into the corona. We simulate how flare-accelerated charged particles propagate down one-dimensional flux tubes and heat the stellar atmosphere using the Fokker-Planck kinetic theory. Detailed radiative transfer is included so that model predictions can be directly compared with observations. The flux of flare-accelerated particles drives return currents which additionally heat the stellar atmosphere. These effects are also included in our models. We examine the impact of the flare-accelerated particle beams on model solar and dMe stellar atmospheres and perform parameter studies varying the injected particle energy spectra. We find the atmospheric response is strongly dependent on the accelerated particle cutoff energy and spectral index.
\end{abstract}

\section{Introduction}\label{sec:intro}
Stellar flares are the result of large explosions in the atmospheres of stars. They are produced when magnetic fields, which have been stressed by convective motion in the stellar photosphere, reconnect rapidly releasing their stored energy. In addition to directly heating the reconnection site in the corona, much of this energy goes into accelerating charged particles to high velocity. These travel down magnetic field lines colliding with the increasingly dense plasma, depositing their energy and momentum, and quickly heating the plasma in the reconnecting flux tubes to temperatures $> 20$ MK. The high temperature causes emission to dramatically brighten in virtually all regions of the electromagnetic spectrum. Flares are ubiquitous and have been observed on stars of nearly all spectral types. 

Because of the Sun's close proximity, understanding solar flares is especially important. Together with coronal mass ejections, these directly affect the Earth environment. They have significant impact on spaced-based communications, the power grid, and the manned space program. Also because of the Sun's proximity, it is possible to spatially resolve many of the features which drive solar flares. Such observations are not possible on other stars. However, despite the lack of spatial resolution, there is still much to be learned by studying stellar flares. (In this paper, we will refer to solar flares as those exclusively on the Sun. Stellar flares will refer to flares on all stars \textit{except} the Sun.) Active dMe stars are known to have flares much larger than those on the Sun. These stars spin faster and generate much larger magnetic fields \citep[and references therein]{2014ApJ...797..121H} resulting in flares some $10^3$ times more energetic than typical solar flares \citep[e.g.,][]{1991ApJ...378..725H, 2003ApJ...597..535H, 2010ApJ...714L..98K}. In addition, the background intensity of dMe stars is much smaller than that of the Sun. Thus, when these stars flare, the signal can be much higher. For example, \citet{2010ApJ...714L..98K} observed the optical luminosity of the dMe star, YZ CMi, to brighten by $\sim 200$ times. In comparison, in the largest solar flare an increase in irradiance of just $\sim 100$ ppm was observed \citep{2006JGRA..11110S14W, 2014ApJ...787...32M}.

Accelerated electrons are known to play an important role in transporting energy during flares. Their presence can be detected from the bremsstrahlung radiation they produce as they collide with the ambient plasma. Since these electrons are a major source of flare heating, it is crucial that we accurately model them. Fortunately, the Ramaty High Energy Solar Spectroscopic Imager (RHESSI; \citealt{2002SoPh..210....3L}) observes this bremsstrahlung radiation from which it is possible to deduce the injected electron spectrum \citep[e.g.,][]{2003ApJ...595L..97H}. Thus, to simulate how the lower atmosphere responds to this heating, we model the precipitation of these electrons from the acceleration site in the corona to the footpoints in the chromosphere and below. 

In addition to electrons, ions are also likely accelerated by reconnecting magnetic fields. \citet{2012ApJ...759...71E} estimated the energy in accelerated ions to be comparable to that of accelerated electrons in many flares. However, since the bremsstrahlung cross-section is inversely proportional to the square of the mass of the colliding particle \citep{1997A&A...326..417H}, their presence is much harder to directly detect. Even with this limitation, RHESSI has detected the presence of accelerated ions in several large flares \citep[e.g.,][]{2006ApJ...644L..93H, 2012ApJ...759...71E}. Even though direct evidence of their presence is scarce, models of particle acceleration predict that lower energy ions ($ < 1$ MeV) will be accelerated --albeit on longer time scales than electrons \citep{2004ApJ...610..550P}. Therefore, to accurately model the flaring atmosphere the effects of flare-accelerated ions must also be included.

A beam of charged particles propagating down a magnetic tube induces an electric field which drives a return current \citep[and references therein]{1990A&A...234..496V, 2006ApJ...651..553Z}. The magnitude of the current depends upon the particle beam spectrum. In fact, the return current alters the particle spectrum causing a flattening at low energies \citep{Holman2012}. This current additionally heats the ambient plasma through Joule heating. In the flaring corona, this can be a major source of energy and could explain superhot coronal temperatures ($> 30$ MK) observed in several large flares \citep[e.g.][]{2010ApJ...725L.161C, 2014ApJ...781...43C}.

It is the purpose of this paper to present a unified computational model which can describe the atmosphere of both solar and stellar flares including the processes most important to flare dynamics. To do this we simulate the transport of a beam of non-thermal particles injected at the top of a magnetic flux tube and follow the subsequent heating of the stellar atmosphere. Our model flux tube extends from the sub-photosphere into the corona. Pressure, temperature, and density vary by many orders of magnitude across these regions, and our models must be able to accurately represent these very different conditions. For example, in the corona radiative transfer is dominated by numerous high temperature, non-LTE, optically thin atomic transitions. In the photosphere, however, radiative transfer is optically-thick and in LTE. In the chromosphere, neither of those approximations hold and the full radiative transfer equation must be solved in detail for several important atomic transitions. Flares produce high-speed shocks (i.e., $> 600$ km s$^{-1}$). We have found that to resolve these requires a computational grid with spacing  $< 100$ m. Currently, models which include detailed radiative transfer at such high-resolution are only tractable in one-dimension. 

Decades of observations have revealed that flares certainly have a three-dimensional geometry. However, the strong magnetic forces present in flaring active regions confine charged particles to flow along field lines. Thus, to a good approximation flare dynamics can be modeled using a one-dimensional geometry with that dimension being the axis of a model flux tube. To partially account for the three-dimensional nature of particular flares, the emission from individual flux tube models can be combined using timing and spatial information provided by observations (e.g., images from the Atmospheric Imaging Assembly on the Solar Dynamics Observatory (SDO/AIA) and RHESSI). In this way, important processes are resolved in ways that are currently not tractable in fully 3D simulations. However, this method does not account for the interaction between plasma on differing flux tubes. This could affect heating rates since those depend on the plasma density which can be sensitive to mixing of plasma on differing tubes. We speculate that the effect of mixing on plasma density will be small compared to that produced by chromospheric evaporation, which is captured by our 1D simulations. This is because neighboring flux tubes are likely to have been heated by similar fluxes of non-thermal particles, so will have similarly elevated densities. In this paper, we present results from parameter studies varying 1D model flux tubes and injected non-thermal spectra. In subsequent papers, we will combine these 1D models to form a more complete representation of particular flares. 

In Section~\ref{sec:radyn}, we describe our method to solve the equations of radiation hydrodynamics and the modeling framework that we use to simulate solar and stellar flares. We discuss how thermal conduction and radiative transfer are included in the models. The chromospheric radiative transfer is of particular importance, and we describe our method to model emission from numerous optically-thick, non-LTE atomic transitions which dominate that region. We also describe how radiative backwarming from coronally produced X-rays and extreme ultraviolet (XEUV) radiation is included. In Section~\ref{sec:beamheat}, we present how our models simulate the precipitation of flare-accelerated particle beams. We present our method for modeling the direct collisional excitation and ionization of the ambient plasma by these particles in Section~\ref{sec:collrates}. In the region of beam impact, these dominate over the thermal rates, significantly altering the radiative transfer. Section~\ref{sec:rc} describes our method for simulating how return currents additionally heat flaring flux tubes. In Section~\ref{sec:parameterstudy}, we present results of a parameter study which we have conducted to understand the range in dynamics predicted from various injected particle beams. In Section~\ref{sec:conc}, we summarize our results and present conclusions.

\section{Radiation Hydrodynamics}\label{sec:radyn}
Flares produce high-speed shocks and increase density and radiation throughout the stellar atmosphere. To model them, the radiative transport equation must be coupled with the standard equations of hydrodynamics. During flares, chromospheric plasma evaporates into the corona and waves can propagate into the photosphere. Modeling the atmospheric response to flare heating requires a model that can extend from the sub-photosphere through the chromosphere, transition region and into the corona. The transition region is extremely narrow, and we have found that accurately resolving it requires spatial scales as small as $\sim100$~m. A model that includes all of these elements represents a major computational undertaking. We use the RADYN code developed by \citet{1992ApJ...397L..59C, 1995ApJ...440L..29C, 1997ApJ...481..500C} to solve the equations of charge and population conservation coupled to the equations of radiation hydrodynamics in 1D. We briefly summarize the method of solution here. The equations of radiation hydrodynamics are, 
\begin{equation}
\pdm{\rho}{t} + \pdm{\rho v}{z}  =  0,
\label{eqn:mass}
\end{equation}
\begin{equation}
\pdm{\rho v}{t} + \pdm{\rho v^2}{z} + \pdm{\left(p + q_v\right)}{z} + \rho g - A_{beam} =  0,
\label{eqn:momentum}
\end{equation}
\begin{align}
& \pdm{\rho e}{t} + \pdm{\rho v e}{z} + (p + q_v) \pdm{v}{z} \nonumber \\ 
& + \pdm{\left(F_c + F_r\right)}{z} - Q_{cor} - Q_{beam}  - Q_{rc} =  0,
\label{eqn:energy}
\end{align}
\begin{equation}
\pdm{n_i}{t} + \pdm{n_i v}{z} - \left(\sum\limits_{j \neq i}^{N'} n_j P_{ji} - n_i \sum\limits_{j \neq i}^{N'} P_{ij} \right)  =  0,
\label{eqn:atompop}
\end{equation}
\begin{equation}
\mu \pdm{I_{\nu \mu}}{z} = \eta_{\nu \mu} - \chi_{\nu \mu} I_{\nu \mu}
\label{eqn:radtrans}
\end{equation}
where $z$, $\rho$, $e$, $v$, and $p$ are the height, density, internal energy density, velocity and pressure, respectively. $g$ is the gravitational acceleration, and $q_v$ is a viscous stress term added to aid in achieving numerical stability. $F_c$ and $F_r$ are the conductive and radiative fluxes, respectively. The conductive flux has the classical Spitzer form but is limited so that it does not exceed the saturation limit of \citet{1980ApJ...238.1126S}. In the atomic level population equation, $n_i$ is the number density in a given atomic state,  and $N'$ is the total number of atomic states considered in these simulations.  We have derived a parameter, $m_0$, which represents the average mass of the plasma per hydrogen atom. This parameter assumes constant abundance ratios using the abundances derived by \citet{2009ARA&A..47..481A}. It has a value of $2.26 \times 10^{-24}$ g.  Thus, $\rho = m_0 n_H$, where $n_H$ is the hydrogen number density. $P_{ij}$ is the transition rate from state $i$ to state $j$ and is given by $P_{ij} = C_{ij} + R_{ij}$, where $C_{ij}$ and $R_{ij}$ are the collisional and radiative rates, respectively. These are discussed in much more detail in \citet[Chapter~5]{Mihalas1978}. In the radiative transfer equation, $I_{\nu \mu}$ is the frequency ($\nu$) and angle ($\mu$) dependent specific intensity, and $\eta_{\nu \mu}$ and $\chi_{\nu \mu}$ are the emission and absorption coefficients, respectively. The radiative flux divergence, $\partial F_r/ \partial z$, is obtained by integrating Eq.~\ref{eqn:radtrans} over frequency and angle. $Q_{cor}$ is a coronal heating term which is necessary to maintain a hot corona. $Q_{beam}$ and $A_{beam}$ are terms describing flare-accelerated particle beam heating and momentum deposition, respectively. They are described in more detail in Section~\ref{sec:beamheat}. $Q_{rc}$ is a heating term due to return currents and is described in Section~\ref{sec:rc}.

These coupled non-linear equations are solved implicitly using a Newton-Raphson iteration scheme. Advected quantities are treated using the second-order upwind technique of \citet{1977JCoPh..23..276V}. RADYN employs an adaptive grid \citep{1987JCoPh..69..175D} which is designed to resolve shocks and steep gradients which can form in flaring stellar atmospheres. The grid cell concentration is chosen to be proportional to the desired resolution.  \citet{1987JCoPh..69..175D} define a resolution operator which is strongly dependent on the absolute value of the gradients of the radiative hydrodynamic variables. Because gradients of these variables can be large at different heights, weighting parameters have been chosen to give preference to those variables that require the most grid sensitivity. In these flare simulations, the largest weights were required on temperature, velocity, and atomic level populations, to properly resolve the transition region, strong shocks that form in the loops, and non-LTE population densities which affect the convergence of the radiative transfer equation. Due to the complexity of the physical problem, we parameterize certain terms in our model using previous work. Thus, some uncertainties in our results may be only due to the uncertainties in these external models. 

\subsection{Optically-Thick Radiative Transfer}
The RADYN code solves the radiative transfer equation for the non-LTE conditions that dominate in the chromosphere. RADYN solves the atomic level population equations (Eq.~\ref{eqn:atompop}) for a six-level with continuum hydrogen atom, a nine-level with continuum helium atom, a six-level with continuum \ion{Ca}{2} ion, and a four-level with continuum \ion{Mg}{2} ion. This allows the calculation of numerous transitions which are important to the chromospheric energy balance. Table~\ref{tab:linetrans} lists the line transitions and Table~\ref{tab:conttrans} lists the continuum transitions that we model in detail. For these transitions, the radiative transfer equation is solved for up to 100 frequency points and five angular points providing us with detailed line profiles. 

\begin{deluxetable*}{lcccccccc}
\tabletypesize{\scriptsize}
\tablewidth{0pt}
\tablecaption{Bound-Bound Transitions \label{tab:linetrans}}
\tablecolumns{6}
\tablehead{
\colhead{Atom} & \colhead{$\lambda_{ij}$ (\AA)\tablenotemark{a}} &
\colhead{Transition} & \colhead{Atom} & \colhead{$\lambda_{ij}$
(\AA)} & \colhead{Transition}  }
\startdata
\ion{H}{1} & 1215.67 & Ly$\alpha$ &
   \ion{Ca}{2} & 8662.14 & $3d\,\,^2\!D_{3/2}$ $\leftrightarrow$ $4p\,\,^2\!P^o_{1/2}$ \\
& 1025.73 & Ly$\beta$ &
    & 8498.02 & $3d\,\,^2\!D_{3/2}$ $\leftrightarrow$ $4p\,\,^2\!P^o_{3/2}$ \\    
& 972.52 & Ly$\gamma$ &
    & 8542.09 & $3d\,\,^2\!D_{5/2}$ $\leftrightarrow$ $4p\,\,^2\!P^o_{3/2}$ \\
& 949.74 & Ly$\delta$ &
   \ion{He}{1} & 625.56 & $1s^{2}$ $^1S_{0}$ $\leftrightarrow$ $1s$ $2s$ $^3S_{1}$\\
& 6562.79 & H$\alpha$ &
    & 601.42 & $1s^{2}$ $^1S_{0}$ $\leftrightarrow$ $1s$ $2s$ $^1S_{0}$ \\
& 4861.35 & H$\beta$ &
    & 10830.29 & $1s$ $2s$ $^3S_{1}$ $\leftrightarrow$ $1s$ $2p$ $^{3}P^{0}_{4}$\\
& 4340.47 & H$\gamma$ &
    & 584.33 & $1s^{2}$ $^1S_{0}$ $\leftrightarrow$ $1s$ $2p$ $^{1}P^{0}_{1}$ \\
& 18751.3 & Pa$\alpha$ &
    & 20581.29 & $1s$ $2s$ $^1S_{0}$ $\leftrightarrow$ $1s$ $2p$ $^{1}P^{0}_{1}$\\
& 12818.1 & Pa$\beta$ &
    \ion{He}{2} & 303.79 & $1s$ $^2S_{1/2}$ $\leftrightarrow$ $1s$ $2p$ $^{2}S_{1/2}$ \\
& 40522.8 & Br$\alpha$ &
    & 303.78 & $1s$ $^2S_{1/2}$ $\leftrightarrow$ $2p$ $^{2}P^{0}_{2}$ \\
\ion{Ca}{2} & 3968.47 & H &
   \ion{Mg}{2} & 2802.70 & h \\
  & 3933.66 & K & 
 & 2795.53 & k
\enddata
\tablenotetext{a}{These are vacuum (air) wavelengths for $\lambda_{ij}$ below (above) 2000~\AA.}
\end{deluxetable*}
\begin{deluxetable*}{lcccccccc}
\tabletypesize{\scriptsize}
\tablewidth{0pt}
\tablecaption{Bound-Free Transitions \label{tab:conttrans}}
\tablecolumns{6}
\tablehead{
\colhead{Atom} & \colhead{$\lambda_{ic}$ (\AA)} &
\colhead{Initial State} & \colhead{Atom} &
\colhead{$\lambda_{ic}$ (\AA)} & \colhead{Initial State} }
\startdata 
\ion{H}{1} & 911 & $n=1$ &
   \ion{He}{1} & 504 & $1s^2$ $^{1}S_0$ \\
& 3646 & $n=2$ &
   & 2600 & $1s$ $2s$ $^{3}S_1$ \\
& 8204 & $n=3$ &
   & 3121 & $1s$ $2s$  $^{1}S_0$\\
& 14584 & $n=4$ &
   & 3421 & $1s$ $2p$ $^{3}P^0_4$ \\
& 22787 & $n=5$ &
   & 3679 & $1s$ $2p$ $^{1}P^0_1$ \\
\ion{Ca}{2} & 1044 & $4s\,\,^2\!S_{1/2}$ &
   \ion{He}{2} & 228 & $1s$ $^{2}S_{1/2}$ \\   
& 1218 &  $3d\,\,^2\!D_{3/2}$&
   & 911 & $2s$ $^{2}S_{1/2}$ \\
& 1219 & $3d\,\,^2\!D_{5/2}$ &
   & 911 & $2p$ $^{2}P^{0}_{1/2}$ \\
& 1417 & $4p\,\,^2\!P^o_{1/2}$ &
   \ion{Mg}{2} & 824 & $2p^6$ $3s$ $^2S_{1/2}$ \\
& 1422 & $4p\,\,^2\!P^o_{3/2}$ &
   & 1168 & $2p^6$ $4p$ $^2P^0_{1/2}$ \\
& & & 
   & 1169 & $2p^6$ $4p$ $^2P^0_{3/2}$
\enddata
\end{deluxetable*}

\subsection{Optically-Thin Radiative Transfer}
\begin{figure}
 \ifeapj \epsscale{1.1} \else \epsscale{1.0} \fi
 \plotone{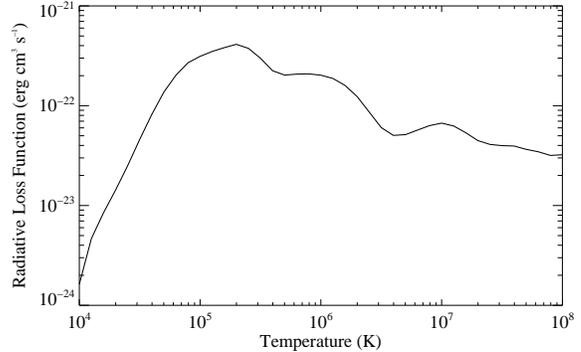}
 \caption{The optically-thin radiative loss function used in these simulations.\label{fig:radloss}}
\end{figure}

The densities in the transition-region and corona are typically low enough that the ``coronal approximation'' applies. In this case, the radiative transfer is dominated by numerous optically thin lines. These lines are formed when ions are collisionally excited and radiatively de-excited. Since the atmosphere is optically-thin in these regions, this radiation is assumed to escape, thus having a net cooling effect on the corona. We model the radiative transfer in these regions using a radiative loss function which is obtained by summing all transitions in the CHIANTI database \citep{1997A&AS..125..149D,2013ApJ...763...86L} except those already accounted for in Tables~\ref{tab:linetrans}~and~\ref{tab:conttrans}. To generate this function the CHIANTI calculations were performed with the assumption of a constant electron density of $10^{10}$ cm$^{-3}$. Figure~\ref{fig:radloss} shows this radiative loss function.

\subsection{XEUV Backwarming}
Half of the optically-thin radiative losses described above are directed outward and leave the stellar atmosphere. The other half are directed downward and will get absorbed in deeper and denser regions. This additional source of heating and ionization in the lower atmosphere becomes especially important during flares when the coronal XEUV emission can become elevated by orders of magnitude. Previously, \citet[hereafter, A05]{2005ApJ...630..573A} developed a method to model how heat is deposited from XEUV backwarming, but that method did not account for the increased photoionizations from this flux. Here we present a new technique which self-consistently includes heating and photoionizations. We have used the CHIANTI database to tabulate emissivities for numerous transitions as a function of temperature and wavelength. RADYN calculates the XEUV spectrum produced from a model loop by integrating the product of these emissivities with the emission measures from the transition region and coronal portions of the loop. Finally, the radiative transfer equation is solved for several optically-thick chromospheric lines, as described above, assuming that this emission is incident on the chromosphere from above. In solving the radiative transfer equation, photoionization cross-sections for XEUV emission are calculated and added to the rates as described by \citet{1994ApJ...433..417W}. 

\begin{figure}
 \ifeapj \epsscale{1.17} \else \epsscale{1.0} \fi
 \plotone{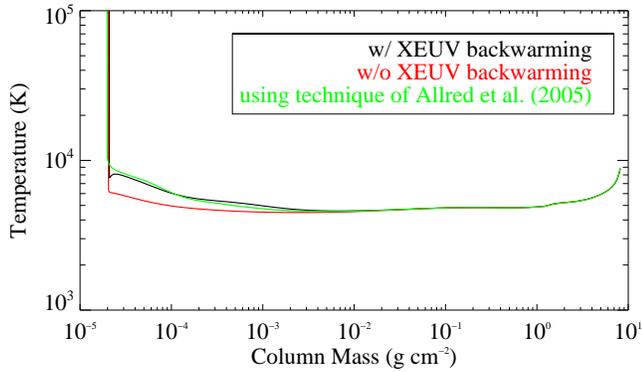}
 \caption{The temperature as a function of column mass for loop models generated using the XEUV backwarming method described here (black), using the XEUV technique of A05 (green) and without XEUV backwarming (red). \label{fig:xeuv}}
\end{figure}

\begin{figure}
 \ifeapj \epsscale{1.17} \else \epsscale{1.0} \fi
 \plotone{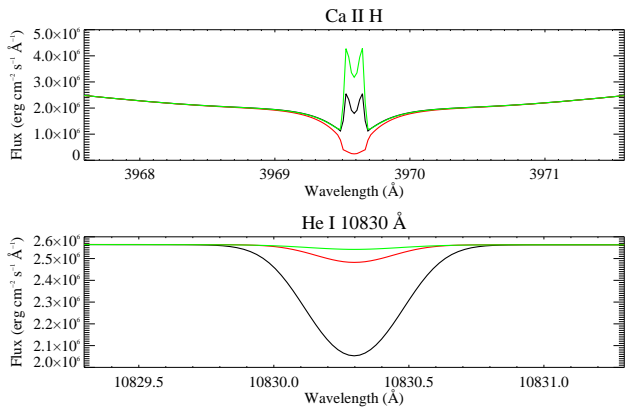}
 \caption{Profiles for the \ion{Ca}{2}~H and \ion{He}{1}~10830~A lines from the loop models generated using XEUV backwarming (black), the technique of A05 (green), and without XEUV backwarming (red). \label{fig:xeuvlines}}
\end{figure}

To understand how XEUV backwarming affects our loop models, we have generated loops which include and exclude this heating term. Figure~\ref{fig:xeuv} compares the structure of the QS.SL.HT loop model (see Section~\ref{sec:initialloops}) which has been generated with and without XEUV backwarming and using the technique of A05. We find that the XEUV backwarming term results in a chromosphere which is 1000 -- 2000 K hotter than would otherwise be expected. The technique presented here and that of A05 produce similar temperature structures. However, the radiative transfer predicted by these methods is significantly different. This is illustrated in Figure~\ref{fig:xeuvlines} which compares the emission from the \ion{Ca}{2}~H and \ion{He}{1}~10830~\AA\  lines for loops generated with and without XEUV backwarming and using the technique of A05. Our technique produces a \ion{He}{1}~10830~\AA\ line with a much deeper absorption profile. This line is formed when continuum photons from the photosphere are absorbed by neutral helium atoms in the $2s$ $^3S_1$ excited state. The XEUV flux increases the photoionization rate of helium. These ions quickly recombine into the $2s$ $^3S_1$ state \citep{2014ApJ...784...30G}, resulting in more absorption than would be expected without the XEUV flux. The \ion{Ca}{2} can be understood similarly. Without the inclusion of the XEUV flux, the region of the chromosphere where the \ion{Ca}{2}~H line center forms is cooler resulting in an overabundance of ground state ions and an overall absorption profile. For both XEUV backwarming techniques, the \ion{Ca}{2}~H line has a central reversal peaking in the near wings. The technique of A05 predicts more flux in the central reversal than our technique. Since their technique does not include direct photoionizations by the XEUV flux, it predicts an overabundance of \ion{Ca}{2} relative to \ion{Ca}{3} ions, and hence an increased flux in the \ion{Ca}{2}~H central reversal.

\subsection{Opacity}
Of course, there are many continuum transitions not included in Table~\ref{tab:conttrans}. The contribution to the opacity due to these transitions is treated using the opacity package of \citet{gustafsson73}. This package constructs opacity as a function of temperature, density and frequency assuming that these transitions are in LTE. These are included as a background opacity source in the detailed calculations of the transitions listed in Table~\ref{tab:conttrans}.

\subsection{Line Broadening}
Observations of flares on dMe stars \citep{1991ApJ...378..725H, 2003ApJ...597..535H, 2010ApJ...714L..98K} and the Sun \citep{JohnsKrull1997} have shown that the hydrogen Balmer lines can become extremely broadened. These authors speculate the cause to be Stark broadening, and that conclusion is supported by the models of \citet{2006ApJ...644..484A} and \citet{2006PASP..118..227P}. In order to test this against other possible sources of broadening, such as thermal or turbulent broadening, we have implemented a technique in RADYN to model Stark line broadening. The amount of Stark broadening depends upon the local electron density. Therefore, determining it can further constrain models of flaring stellar atmospheres.

That the orbital angular momentum states of hydrogen labeled by the quantum number, $l$, are not degenerate in the presence of an electric field perturbation is described by the well-known Stark effect \citep[e.g.][]{Kepple1968, Vidal1971, Seaton1990}. In stellar atmospheres, this perturbation is due to a net electric microfield from fluctuations in the ambient electron, proton, and ion density. The first order (linear) energy level shifts are proportional to the electric microfield strength \citep[which has a probability distribution typically modeled as a Holtsmark or Hooper distribution;][]{Nayfonov1999}, the principal quantum number, $n$, and a quantum number, $q$, which describes the relationship between the parabolic quantum numbers $q_1$ and $q_2$ where $q = q_1 -q_2$ and $q_1 + q_2 = n - \mid m \mid - 1$ \citep{Condon1935}. Therefore, the higher order hydrogen lines within a series experience a larger amount of Stark broadening.

The general prescription for line cross-section is given by a Voigt profile with a damping parameter, $\Gamma$, that includes separate terms for radiative damping, $\Gamma_{\mathrm{rad}}$, and for collisions among neutrals (resonance and van der Waals), $\Gamma_{\mathrm{res/vdW}}$. A microturbulence of 2~km~s$^{-1}$ is included in the Doppler width. The Voigt profile is then convolved with the Stark profile to give the total line cross-section \citep{Mihalas1978}. However, as an approximation to the more complete treatment, the Voigt profile damping parameter, $\Gamma$, can be modified to also include a Stark damping term, $\Gamma_{\mathrm{s}}$, such that $\Gamma = \Gamma_{\mathrm{rad}} + \Gamma_{\mathrm{res/vdW}} + \Gamma_{\mathrm{s}}$. This has been found to work well for solar hydrogen lines in the infrared \citep{CarlssonRutten1992}, so we have chosen this method for modeling Stark broadening in RADYN.

The Stark damping parameter for hydrogen is taken from the approximate line shape formulae in Method~\#1 of \citet{Sutton1978}. In using this method, we are consistent with other NLTE radiative transfer codes such as RH \citep{Uitenbroek2001}. The Stark damping profile is given by
\begin{equation} \label{eq:sutton}
\Gamma_s = 0.6 g_q (j^2 - i^2)  n_e^{2/3}
\end{equation}
where $g_q=0.642$ for the $\alpha$ transition of each series and $g_q=1$ otherwise, and $j$ and $i$ are the upper and lower levels of the transition, respectively. For transitions of helium, calcium, and magnesium, $\Gamma_s = q_s n_e$, where $g_s$ is a constant resulting in Stark broadening which is approximately three orders of magnitude less than for hydrogen.

\subsection{Boundary Conditions}
Our model flux tubes are assumed to be symmetric about the loop apex, so that we need only model one half of a full loop. The top boundary is at the loop apex where we have implemented a reflecting boundary condition to mimic incoming waves from the other side of the loop. The bottom boundary is below the photosphere where densities and pressures are very high. There we have implemented a simple transmitting boundary to allow any waves which reach that level to pass through into the interior.

\subsection{Initial Loops}\label{sec:initialloops}
A focus of this work is to study how particle beams, which are known to be accelerated during flares, deposit their energy in magnetic flux loops in stellar and the solar atmospheres. To this end, we have generated several initial loop states with diverse lengths, temperatures, and densities which we will use in this study. These extend from the sub-photosphere through the corona. The corona is kept hot by adding a heating term, $Q_{cor}$, which is chosen to just balance the conductive and radiative losses in the upper coronal portion of the loop. Heat is also added to the sub-photosphere to balance radiative and convective flux losses there. With these heat sources, the loops are allowed to relax until a state of near equilibrium is reached. 

\begin{figure*}
 \ifeapj \hspace*{-.4in} \epsscale{1.2} \else \hspace*{-.4in} \epsscale{1.16} \fi
\plottwo{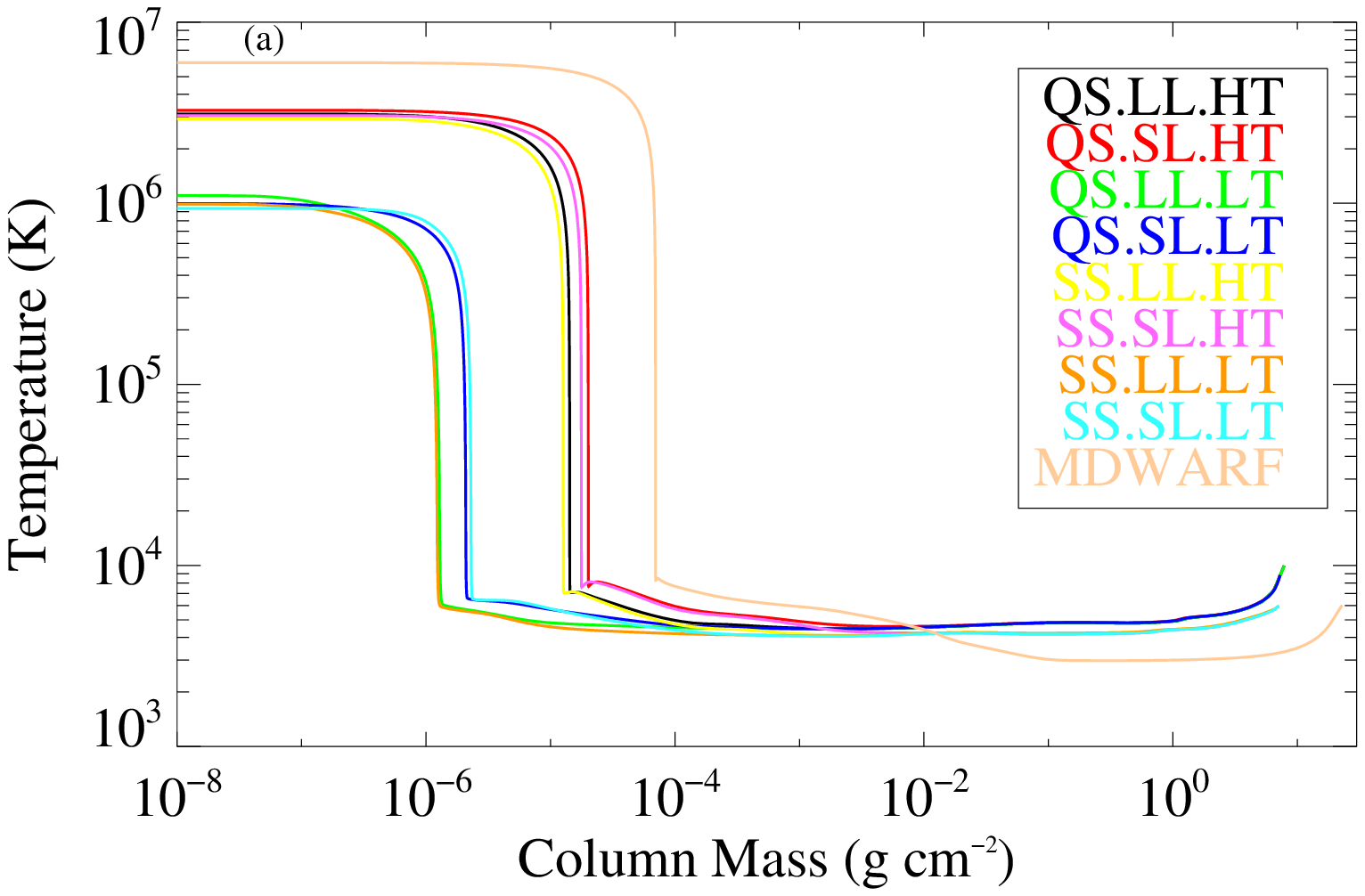}{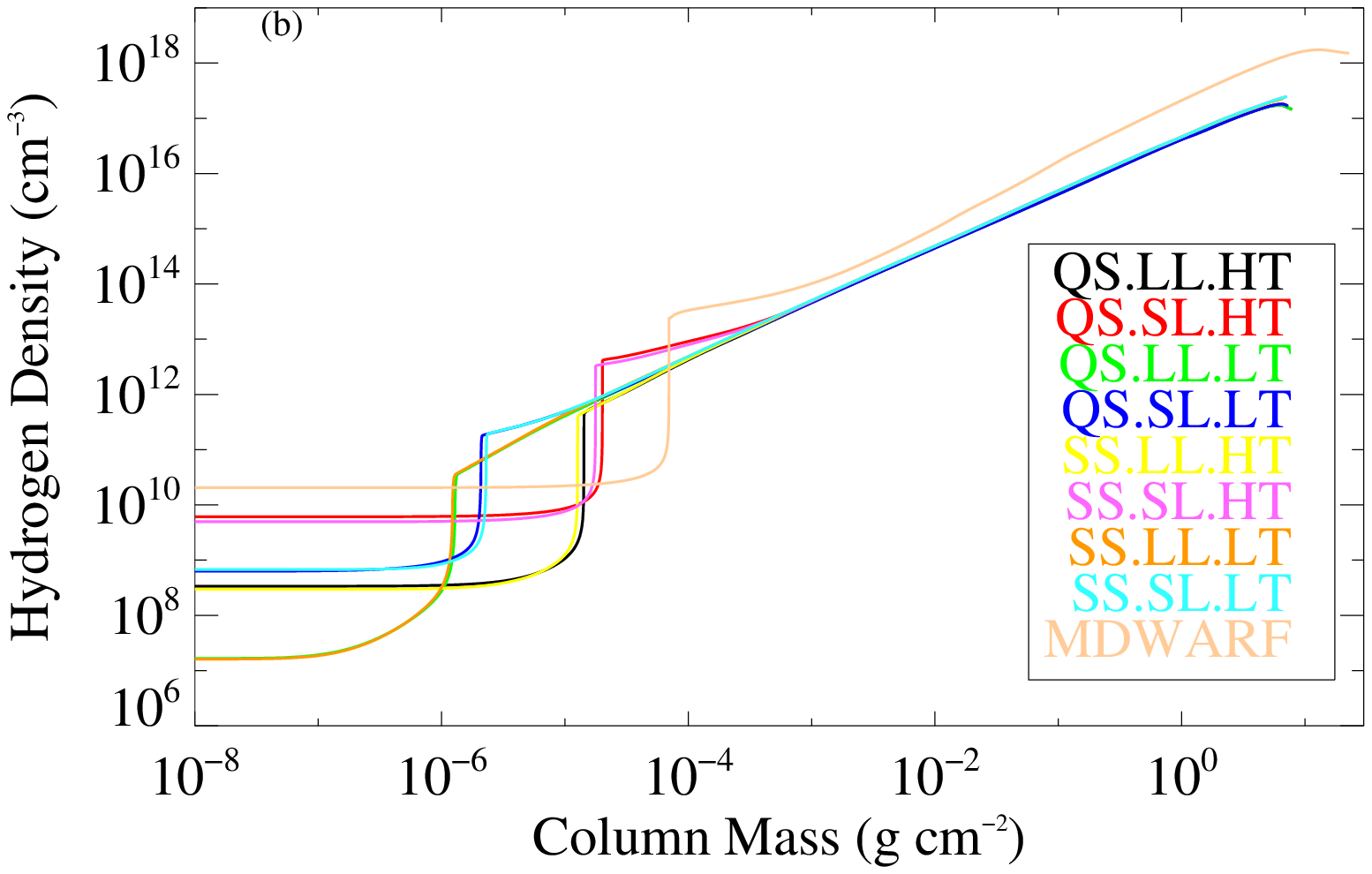}
\caption{The temperature (a) and hydrogen number density (b) as a function of column mass for the loops described in Table~\ref{tab:initial_loops}. \label{fig:initial_loops}}
\end{figure*}

We have generated solar-type initial loops using three free parameters. These are the photospheric temperature, loop-length, and coronal temperature. The coronal density is dependent on loop-length and coronal temperature by an RTV-type \citep{1978ApJ...220..643R} scaling law so cannot be independently varied. We have chosen photospheric temperatures (i.e., the temperature in our loops where $\tau_{5000}  = 1$) of 5000~K and 5800~K which correspond with sunspot and ``quiet Sun'' conditions. Loop-lengths were chosen to be 10 Mm for short loops and 100 Mm for long loops. So that we can model flaring flux tubes in dMe stellar atmospheres, we have also produced a model loop appropriate for M dwarf atmospheres. This loop has a higher surface gravity ($562$~m~s$^{-2}$), cooler photosphere (3500~K), and hotter corona (6 MK) than the solar loops. The parameters describing these loops are summarized in Table~\ref{tab:initial_loops} and the temperature and density of the loops are plotted in Figure~\ref{fig:initial_loops}. 

As described in Section~\ref{sec:beamheat}, the energy deposition rate from particle beams depends upon the magnetic field in the loops. To account for this we assume a magnetic field strength in the photosphere of 1 kG for solar loops and 5 kG for the M dwarf loop. Spatially-averaged observations of magnetic fields on active M dwarf stars have found that the majority of the stellar surface ($60 - 70\%$) has a magnetic field strength of $3 - 4$ kG \citep{1985ApJ...299L..47S,1994IAUS..154..493S,1996ApJ...459L..95J}. Since these are spatial averages and large flares are likely to occur where the field is strongest, we have estimated a photospheric field of 5 kG in the loop. At the loop tops, the magnetic field strength is assumed to be 100 G for the shorter (10 Mm) solar loops, 10 G for the longer (100 Mm) solar loops, and 500 G for the M dwarf loop. These values were chosen to be consistent with active region coronal magnetic field extrapolations \citep[e.g.][]{2012SoPh..281...53T}. The magnetic field in all cases is assumed to exponentially decrease between these boundary conditions.
\begin{deluxetable*}{lcccc}
 \tabletypesize{\scriptsize}
 \tablewidth{0pt}
 \tablecaption{Initial Loops\label{tab:initial_loops}}
 \tablecolumns{5}
 \tablehead{
 \colhead{Label} & \colhead{Photospheric Temperature (K)} & \colhead{Loop-length (Mm)} & \colhead{Coronal Electron Density (cm$^{-3}$)} & \colhead{Coronal Temperature (MK)}  }
\startdata
QS.LL.HT & 5800 & 100 & $\sn{3}{8}$ & 3.0 \\
QS.LL.LT & 5800 & 100 & $\sn{2}{7}$ & 1.0 \\
QS.SL.HT & 5800 & 10 & $\sn{6}{9}$ & 3.0 \\
QS.SL.LT & 5800 & 10 & $\sn{6}{8}$ & 1.0 \\
SS.LL.HT & 5000 & 100 & $\sn{3}{8}$ & 3.0\\
SS.LL.LT & 5000 & 100 & $\sn{2}{7}$ & 1.0\\
SS.SL.HT & 5000 & 10 & $\sn{6}{9}$ & 3.0 \\
SS.SL.LT & 5000 & 10 & $\sn{6}{8}$ & 1.0 \\
M dwarf & 3500 & 10 & $\sn{2}{10}$ & 6.0 
\enddata
\end{deluxetable*}

\section{Particle Beam Heating}\label{sec:beamheat}
Accelerated particle beams are the main source of heating in the lower atmosphere during flares, so it is crucial that we accurately model their propagation and energy deposition as they move through magnetic flux loops. The rate of energy lost by a particle beam depends upon the details of the beam's energy spectrum as well as the composition and ionization states of the ambient plasma. We use the Fokker-Planck method coupled with results from our radiative hydrodynamic simulation described in Section~\ref{sec:radyn} to determine the distribution function, $f(\vect{x}, \vect{p}, t)$, for flare-accelerated particles. Because the magnetic field constrains the beam particles to move along field lines, we can replace the vector position, $\vect{x}$, with the coordinate, $z$, which measures the distance along the field line. We replace the vector momentum, $\vect{p}$, with the kinetic energy, $E$, and particle pitch angle, $\alpha$. In the following discussion, we will simply refer to $\mu = \cos(\alpha)$ as the pitch angle. The transit time for beam particles from injection source to footpoints is fast relative to changes in the beam particle energy distribution and hydrodynamic time scales so a time-independent distribution function is assumed. Thus, we will solve the kinetic equation for the distribution, $f(E,\mu, z)$, where $f(E,\mu, z) \textrm{d}E \textrm{d}\mu \textrm{d}z$ is the number density of beam particles with energy between $E$ and $\textrm{d}E$, pitch angle between $\mu$ and $\textrm{d}\mu$, and position between $z$ and $\textrm{d}z$. In solving the kinetic equation, we include the effects of particles moving at relativistic speeds, Coulomb collisions, synchrotron emission, pitch angle scattering, and magnetic field gradients. However, we neglect effects due to plasma turbulence, electric fields and self-absorption of synchrotron emission. We follow \citet{1990ApJ...359..524M} and write the Fokker-Planck equation as 
\begin{align}\label{eqn:fp}
& \mu \pdm{\Phi}{z} - \frac{d \ln B}{2 dz} \pdm{}{\mu} \left[ \left( 1 - \mu^2 \right)\Phi \right]  =  \nonumber \\
& \frac{1}{\beta^2} \pdm{}{E}\left\{ \left[C + S \beta^3 \gamma^2 \left( 1 - \mu^2 \right) \right] \Phi \right\}  \nonumber \\
& - \frac{S}{\beta \gamma} \pdm{}{\mu} \left[ \mu \left( 1 - \mu^2 \right) \Phi \right] \nonumber \\
& + \frac{C'}{\beta^4 \gamma^2} \pdm{}{\mu} \left[ \left( 1 - \mu^2 \right) \pdm{\Phi}{\mu} \right]
+ \frac{\Sigma}{c \beta^2}
\end{align}
where $\Phi = f/\beta$, $\gamma = E + 1$ is the relativistic total energy with $E$ measured in units of $m c^2$, $\beta c$ is the relativistic ion velocity, $v$, $B$ is the magnetic field strength in the loop, and $\Sigma$ is a source term for particles injected at the loop top. $C$ and $C'$ are coefficients which measure the beam energy loss rate and pitch angle diffusion from collisions, respectively. Similarly, $S$ measures the loss rate and pitch angle diffusion due to synchrotron emission. \citet{1990ApJ...359..524M} developed a numerical method to solve this equation for flare-accelerated electrons. We have generalized their method to solve the Fokker-Planck equation for both flare-accelerated electrons and ions. This has required altering $C$, $C'$, and $S$ to account for the physics which characterize ion beams. Here we summarize this theory.

\citet{1985ApJ...299..987P} calculated the energy loss and pitch angle scattering rates due to synchrotron radiation. From those results, we find the synchrotron coefficient, $S$, for both ion and electron beams to be \begin{equation}
 S = \frac{2}{3}\frac{\left(Z e\right)^4 B^2}{\left(m c^2\right)^3}
\end{equation}
where $e$ is the proton charge, and $Z$ is the number of protons (or electrons) in the beam particle. 

$C$ is related to the energy loss due to collisions by
\begin{equation} \label{eqn:c}
C = -\frac{dE_{col}}{dt} \frac{\beta}{c}
\end{equation}
\citet{1990ApJ...359..524M} assumed that the ambient plasma was a ``cold target'' meaning that the beam particle's velocity is much greater than the thermal velocity for all constituents of the ambient plasma. Formally, this is expressed as $x_i \gg 1$ where $x_i =\left( v/v_i \right)^2 = (m_i/m) E/(k T_i)$ and $m_i$, $v_i$, and $T_i$ are the mass, velocity and temperature for plasma constituent,~$i$, and $m$ is the mass of the beam particle. For flare-accelerated electrons even at the low energy limit of $\sim 10$ keV, the ambient electron temperature would have to be greater than 100 MK for the cold target approximation to fail. This far exceeds observed temperatures for even the largest flares. Therefore, beam electrons can always be treated using cold target collision theory. With ion beams, however, this is not necessarily true. Flare-accelerated protons of $\sim 1$ MeV (which is about the low energy detection limit of current instruments) have a velocity less than that of thermal electrons at temperature $\gtrsim 6$~MK. Flares routinely produce plasma much hotter than this. Therefore when modeling how ion beams interact with the ambient plasma, we must employ a theory which can account for both hot and cold targets. \citet{1965RvPP....1..105T} developed this theory and found the energy loss rate of particles with energy, $E$, from collisions with ambient charged particles of species, $i$, to be 
\begin{equation} \label{eqn:dedtcharged}
\frac{dE_i}{dt} = -E \nu_i
\end{equation}
where
\begin{equation}
\nu_i = 2 \nu_{0_i} \frac{m}{m_i} \left[ Erf \left(\sqrt{x_i} \right) - \frac{2}{\sqrt{\pi}} \sqrt{x_i} e^{-x_i} 
\left(1+\frac{m_i}{m} \right) \right]
\end{equation}
and
\begin{equation}
\nu_{0_i} = \frac{2 \pi e^4 Z^2 Z_i^2 \lambda_i n_i }{E m^2 c^2 v}
\end{equation}
$Z_i$ is the target particle charge in units of $e$, and $n_i$ is the number density for target particle, $i$. $\lambda_i$ is the Coulomb logarithm, and is defined by $\lambda_i = \ln(r_{max}/r_{min})$. In this case $r_{max}$ is the mean free path length $\eta = v/\nu$ \citep[hereafter, referred to as E78]{1978ApJ...224..241E} where $\nu$ is the plasma oscillation frequency. $r_{min}$ is given by $r_{min} = \hbar / 2 M_i v$ where $M_i$ is the reduced mass of the beam and target particle \citep{huba11}. Combining these gives 
\begin{equation}
\lambda_i =  \ln \left( \frac{M_i v^2}{\hbar} \left[ \frac{m_i}{\pi n_i Z_i^2 e^2} \right]^{\frac{1}{2}}\right)
\end{equation}

The collisional pitch angle diffusion coefficient, $C'$, is obtained from the rate that beam particles are scattered out of the direction of beam propagation. It is given by 
\begin{equation}
C' = \frac{\beta^2}{c^2} \frac{dv}{dt}\label{eqn:cprime}
\end{equation}

For collisions with neutral targets (e.g., neutral hydrogen or helium), the beam particles interact with atomically bound electrons. As long as the beam velocity is much greater than the Bohr orbital velocity, neutral atoms can be treated as a cold targets. Using the fine structure constant to approximate the orbital velocity, we find that the cold target approximation for collisions with neutral hydrogen is good for protons with energy $\geq 25$ keV and electrons $\geq 0.01$ keV. The energy loss rate for a charged beam on a neutral target is given by \citep[E78;][p.~614]{mott65} 
\begin{equation} \label{eqn:dedtneutral}
\frac{dE_n}{dt} = \frac{-2 \pi e^4}{E} \frac{m}{m_e} Z^2 Z_n n_n \lambda_n v 
\end{equation}
where $n_n$ is the number density for the ambient neutral atom, $n$. $Z_n$ is the number of electrons in the atom, and $\lambda_n$ is an effective Coulomb logarithm for collisions with the atom. For an atom with mean ionization energy, $I_n$, $\lambda_n$ is given by \citep[p.~637]{evans55}
\begin{equation}
\lambda_n = \ln \left( \frac{2 m_e v^2 \gamma^2}{I_n} \right) 
\end{equation}

The total energy loss rate due to collisions is given by 
\begin{equation} \label{eqn:dedttotal}
\frac{dE_{col}}{dt} = \displaystyle\sum_i \frac{dE_i}{dt} + \displaystyle\sum_n \frac{dE_n}{dt} 
\end{equation}
where the sums are over charged and neutral species, respectively. In calculating the beam energy deposition rate in RADYN, we consider a plasma consisting of electrons, protons, neutral hydrogen, neutral, singly, and doubly ionized helium, and singly ionized calcium and magnesium. These are the species most important to the energetics of the chromosphere where the majority of the beam impacts. However, since the energy loss rates are inversely proportional to the target particle's mass, collisions with electrons --both ambient and those in neutral atoms-- dominate the energy transfer. We use Equation~\ref{eqn:dedtcharged} for calculating the energy loss rate for collisions with charged particles and ions, and Equation~\ref{eqn:dedtneutral} for collisions with neutral particles. We use Equations~\ref{eqn:dedttotal}~and~\ref{eqn:c} to obtain $C$, which is needed to solve Equation~\ref{eqn:fp}. The number densities for each of these species is calculated in detail in the simulations. Thus, the beam energy deposition drives the simulations which alter the beam deposition in a self-consistent way.

By solving Equation~\ref{eqn:fp}, we obtain the distribution function, $f$, from which we can obtain the particle beam heating rate ($Q_{beam}$) and momentum deposition rate ($A_{beam}$) using the following relations:
\begin{equation}
Q_{beam} = \frac{\mathrm d}{\mathrm d z} \left( \int_{\mu} \! \int_{E} \! \mu v E f \, \mathrm{d}E \mathrm{d}\mu \right)
\end{equation}
and
\begin{equation}
A_{beam} = \frac{\mathrm d}{\mathrm d z} \left( \int_{\mu} \! \int_{E} \! \mu v a f \, \mathrm{d}E \mathrm{d}\mu \right)
\end{equation}
where $a = \gamma m v$ is the relativistic momentum. Even though we have developed and presented here a method to simulate both flare-accelerated electrons and ions, in the remainder of this paper we will focus on the effects of electron beams. A detailed study of the for ion beams will be presented in a future paper.

\section{Particle Beam Collision Rates}\label{sec:collrates}
During flares, the transition rates, $P_{ij}$, can be very enhanced due to direct ionizations and excitations caused by collisions of non-thermal beam particles with the ambient plasma. In this section, we describe a method for incorporating these elevated rates in the most important species (i.e., hydrogen and helium) in RADYN.    

\subsection{Hydrogen} \label{sec:hrates}
The non-thermal collisional rate from a flux of charged particles is given by the general formula \citep[Chapter \textsc{xvi}]{mott65},
\begin{equation}\label{eq:ntrates}
C_{beam}^{nt} = 2\pi \int \int \mu v_B f \sigma_{ij} dE d\mu
\end{equation}
where $f$ is the beam particle distribution function, $\mu$ is the particle pitch angle as described in Section~\ref{sec:beamheat}, and $\sigma_{ij}$ is the cross section for ionization or excitation. 

Since the flare-accelerated particles have much larger energy than the ionization potential of hydrogen, secondary ionizations from electrons liberated in the primary ionization need to be included. \citet{DalganoGriffing} have performed calculations to include these secondary ionizations, giving an average energy of 36~eV per electron-ion pair produced in the complete absorption of beam particles with energy in the range of $200 - 1000$~eV by a cold hydrogen gas. The amount of energy produced per electron-ion pair is constant for $E>$200~eV, and we assume that this extends to all electron energies above 1000~eV. Including the secondary ionizations, the non-thermal collisional ionization rate can be written in terms of the amount of energy lost by the beam to neutral hydrogen, $\frac{dE_H}{dt}$ obtained from Equation~\ref{eqn:dedtneutral}, \citep[][see also \citealt{Ricchiazzi1983}]{Fang1993},
\begin{equation}
\frac{dE_H}{dt} = 36~\mathrm{eV} \times n_1 C_{beam,1c}^{nt}
\end{equation}  
assuming that most of the neutral atoms are in the ground state, $n_1$. 

A comprehensive study of non-thermal rates in solar flares was presented by \cite{Ricchiazzi1982, Ricchiazzi1983}, who found that the \ion{H}{1} rates of non-thermal collisional ionization from the ground state (denoted, 1-c) and collisional excitation from the ground to first excited state (denoted, 1-2) were important compared to their respective thermal rates, but that the non-thermal 2-c rate was negligible compared to the thermal rate for the applied range of beam fluxes. \cite{Fang1993} has presented revised \ion{H}{1} 1-c, 1-2, 1-3, 1-4 non-thermal ionization and excitation rates, which were adopted by \cite{Kasparova2009} in their study of the hydrogen lines in response to electron beams. Notably, the 1-c rate is a factor $\sim$4.6 higher compared to the \cite{Ricchiazzi1983} 1-c rate. To facilitate comparison with the most recent studies, we use the constants ($R^{H,nt}$) from \cite{Fang1993} to derive the non-thermal collisional rates with hydrogen, according to the formula, 
\begin{equation}
C_{beam}^{nt} = R^{H,nt} \frac{\Lambda_n}{n_e \Lambda_i + n_H \Lambda_n} Q_{\mathrm{beam}}
\label{eqn:hrates}
\end{equation}
where $n_H$ is the neutral hydrogen density and $R^{H,nt}=\sn{1.73}{10}$, $\sn{2.94}{10}$, $\sn{5.35}{9}$, $\sn{1.91}{9}$ for the 1-c, 1-2, 1-3, and 1-4 transitions, respectively.

\subsection{Helium} \label{sec:herates}
A05 found that the ionization fraction of \ion{He}{2} controls how a flaring flux tube transitions from gentle to explosive chromospheric evaporation. To more accurately model these dynamics, we include the non-thermal rates for the primary ionization of neutral helium (\ion{He}{1} 1s$^2$ $\rightarrow$ \ion{He}{2} 1s) and singly ionized helium (\ion{He}{2} 1s  $\rightarrow$ \ion{He}{3}). \citet{Younger1981} calculated parameterized cross sections for these transitions as a function of impacting electron energy. For clarity, we reproduce his result here,
\begin{align}
\sigma_{ij} =  \frac{1}{u I^2} & \left[ A \left(1-\frac{1}{u} \right) + B \left(1-\frac{1}{u} \right)^2 \nonumber \right. \\
             & + \left. C \ln(u) + \frac{D}{u} \ln(u) \vphantom{\left(\frac12^2\right)^2} \right]
 \label{eqn:hecolcs}
\end{align}
where $I$ is the ionization energy and $u = E/I$ is the normalized energy of the colliding electron. $A$, $B$, $C$, and $D$ are constants provided by \citet{Younger1981} and \citet{Arnaud1985} and have values of $17.8$, $-11.0$, $7.0$, $-23.2$ and $14.4$, $-5.6$, $1.9$, $-13.3$ in units of $10^{-14}$ cm$^2$ eV$^2$ for \ion{He}{1} and \ion{He}{2}, respectively. With these cross sections and the beam particle distribution function, $f$, obtained from Equation~\ref{eqn:fp}, we calculate the non-thermal helium ionization rates using Equation~\ref{eq:ntrates}.

\section{Return Current}\label{sec:rc}
Images of the footpoint locations of positively-charged ion beams and negatively-charged electron beams indicate that they are not co-spatial \citep{2006ApJ...644L..93H}. For an electron beam without a co-streaming positively charged beam of equal current density, the charge imbalance between the acceleration region and any given point along the magnetic loop leads to a macroscopic \emph{return current} electric field \citep{Hoyng1976, Knight1977,1990A&A...234..496V} in a direction along the beam. The return current electric field decelerates the beam electrons while accelerating ambient electrons\footnote{Ambient protons are also accelerated, but the drift velocity is much lower due to their larger mass.}, which drift in the direction opposite that of the return current electric field towards the loop top with a Maxwellian distribution of speeds, assuming that the return current field is not super-Dreicer \citep{Holman1985}.  These drifting electrons form the \emph{return current}, which  heats the atmosphere through Ohmic (Joule) dissipation. Many aspects of the beam-return current-atmospheric system have been studied, including pitch angle modifications of the beam \citep{Emslie1980}, return current collisional rates \citep{Karlicky2004}, return current-beam instabilities and subsequent particle  acceleration \citep{Karlicky2012, Pech2014}, and turbulent effects on the beam-return current system \citep{Kontar2014, Xu2013}. The return current modifications on the classical thick target model \citep{Brown1971} have been used to explain the difference between looptop and footpoint hard X-ray spectral indices \citep{Battaglia2008, Xu2013}.  

Recently, \citet{Holman2012} derived return current heating rates in flaring coronal conditions. After the first short moments of beam propagation, a relatively steady-state is quickly attained \citep{1990A&A...234..496V} in which the magnitude of the return current density is equal the beam current density. The return current electric field can be determined from Ohm's law:  $E_{rc}=\eta J_{\mathrm{beam}}$, where $\eta$ is the classical Spitzer resistivity and $J_{\mathrm{beam}}$ is the return current density. \citet{Holman2012} determined the return-current plasma heating rate for a given electron beam flux self-consistently accounting for Joule heating and the reduction of beam electrons due to (non-collisional) thermalization caused by the return current electric field. However, this treatment does not self-consistently account for the change in flux of beam electrons due to Coulomb collisions with the ambient plasma and is therefore an overestimate, especially in the chromosphere where Coulomb collisions are large. In the corona, however, there are relatively few collisions, and this approximation works quite well. Additionally, return current heating is likely largest in the corona, since it is there that beam current densities are highest. Therefore, to account for return current heating in flares, we have chosen to incorporate the formalism of \citet{Holman2012} into RADYN. The result for the volumetric return current heating rate from a power-law electron beam spectrum with power-law index, $\delta$, and cutoff energy, $E_c$, is given by
\begin{align}
Q_{rc}(x) = 
  \begin{cases}
     \eta e^2 {F_e}^2 & x < x_{rc} \\
     \eta e^2 {F_e}^2  \left( \delta \frac{E_{\mathrm{therm}}}{E_c} + \frac{V(x)}{E_c}\right) \\
     \times \left(\frac{E_{\mathrm{therm}}}{E_c} + \frac{V(x)}{E_c}\right)^{1-2\delta} & x \ge x_{rc} 
  \end{cases}
  \label{eqn:rcheat}
\end{align}
where $x$ is distance from the loop top and $F_e$ is the beam particle flux given by $\int \int \mu v f \, \mathrm{d}E \mathrm{d}\mu$. Beam electrons are assumed to be thermalized and removed from the beam when their energy is $E_{\mathrm{therm}} = 2.5 k T$. $x_{rc}$ is the distance at which the lowest energy electrons in the beam are thermalized, and $V(x)$ is the electric potential energy as described in \citet{Holman2012}. A more complete treatment which includes return current losses directly in the Fokker-Planck equation (Eqn.~\ref{eqn:fp}), similar to work done by \citet{Zharkova2005} and \citet{Dobranskis2014}, is outside the scope of this paper but will be presented in a future paper. 

\section{Parameter Study} \label{sec:parameterstudy}
The energy distribution of flare-accelerated particles can vary greatly between particular flares. Since the impact location strongly depends on the particle distribution, that distribution is very important in determining how stellar atmospheres respond to flare heating. To explore the range of possibilities, we have performed a parameter study varying the injected electron beam spectrum, pitch-angle distribution, and the initial loop conditions into which these particles impact.

Flare-accelerated particles are typically observed to have a power-law energy distribution with a cutoff energy, $E_c$, below which relatively few particles are accelerated. Cutoff energies as high as 250 keV have been observed (R. Schwartz, personal communication, 2015). But in many flares, the cutoff energy cannot be directly determined, since thermal X-ray emission in the range below $10 - 20$ keV from the flare-heated plasma overwhelms the non-thermal bremsstrahlung emission. In these flares, it is only possible to determine an upper limit to the cutoff energy. To explore how the atmosphere responds to beams of particles with energies below this upper limit, we have chosen a lower limit of 5 keV. Therefore, to account for the full range in parameter space, we have varied the cutoff energy between $5 - 500$~keV. The slope of the power-law is given by the spectral index, $\delta$. Typical power law indices range from $3 - 9$, but in this study we have varied it between $2.5 - 10$ to ensure that we have spanned the full range of possible values. We have also varied the pitch angle distribution with which the flare-accelerated particles were injected. We modeled this distribution as a Gaussian centered around the flux loop axis. Thus, it has the form, $e^{((\mu-1)/\mu_0)^2}$, where $\mu_0$ is a parameter which controls the width of the Gaussian distribution. In addition to the Gaussian distribution, we have included simulations in which the particles are fully beamed, i.e., they all start with a $0^\circ$ pitch-angle, and simulations with isotropic distributions. We have modeled the effects of beam impact varying these parameters on each of the loops in Table~\ref{tab:initial_loops}. The parameters are summarized in Table~\ref{tab:parameterstudy}. To perform this study, we ran more than $11,000$ simulations.

\begin{deluxetable}{cc}
 \tabletypesize{\scriptsize}
 \tablewidth{0pt}
 \tablecaption{Parameter Study\label{tab:parameterstudy}}
 \tablecolumns{2}
 \tablehead{
 \colhead{Parameter} & \colhead{Range} }
\startdata
$E_c$ & $5 - 500$ keV\\
$\delta$ & $2.5 - 10$\\
$\mu_0$ & $0.1 - 1.0$, isotropic, beamed\\
Loop Conditions & Those listed in Table~\ref{tab:initial_loops}
\enddata
\end{deluxetable}

\subsection{Penetration Depths in Flux Loops}
\begin{figure}
 \ifeapj \hspace*{-.1in} \epsscale{1.18} \else \epsscale{.55} \fi
 \plotone{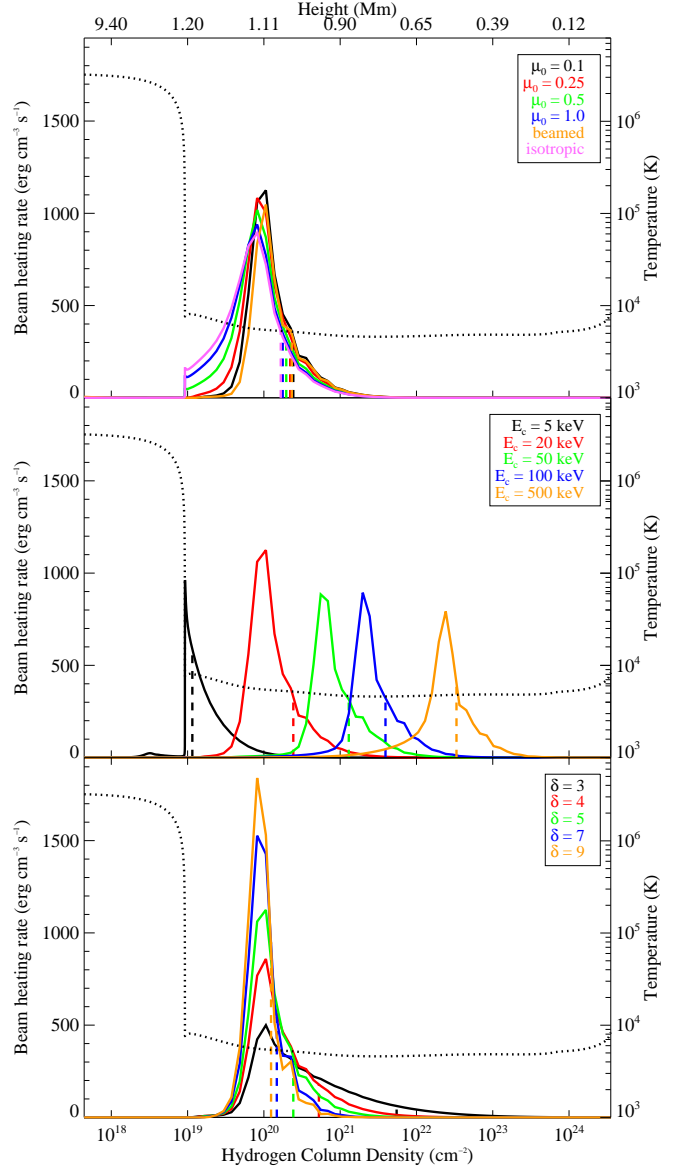}
 \caption{The beam heating rate for several different parameters in the QS.SL.HT loop. In the top panel, $\mu_0$ is varied while holding $E_c = 20$ keV and $\delta = 5$. In the middle panel, $E_c$ is varied while holding $\delta = 5$ and $\mu_0 = 0.1$. In the bottom panel, $\delta$ is varied while holding $E_c = 20$ keV and $\mu_0 = 0.1$. In each panel, the dotted line indicates the temperature structure in the QS.SL.HT loop. \label{fig:avehexamp}}
\end{figure}

\begin{figure*}
 \ifeapj \hspace*{-.55in} \epsscale{1.25} \else \hspace{-.45in} \epsscale{1.16} \fi
 \plottwo{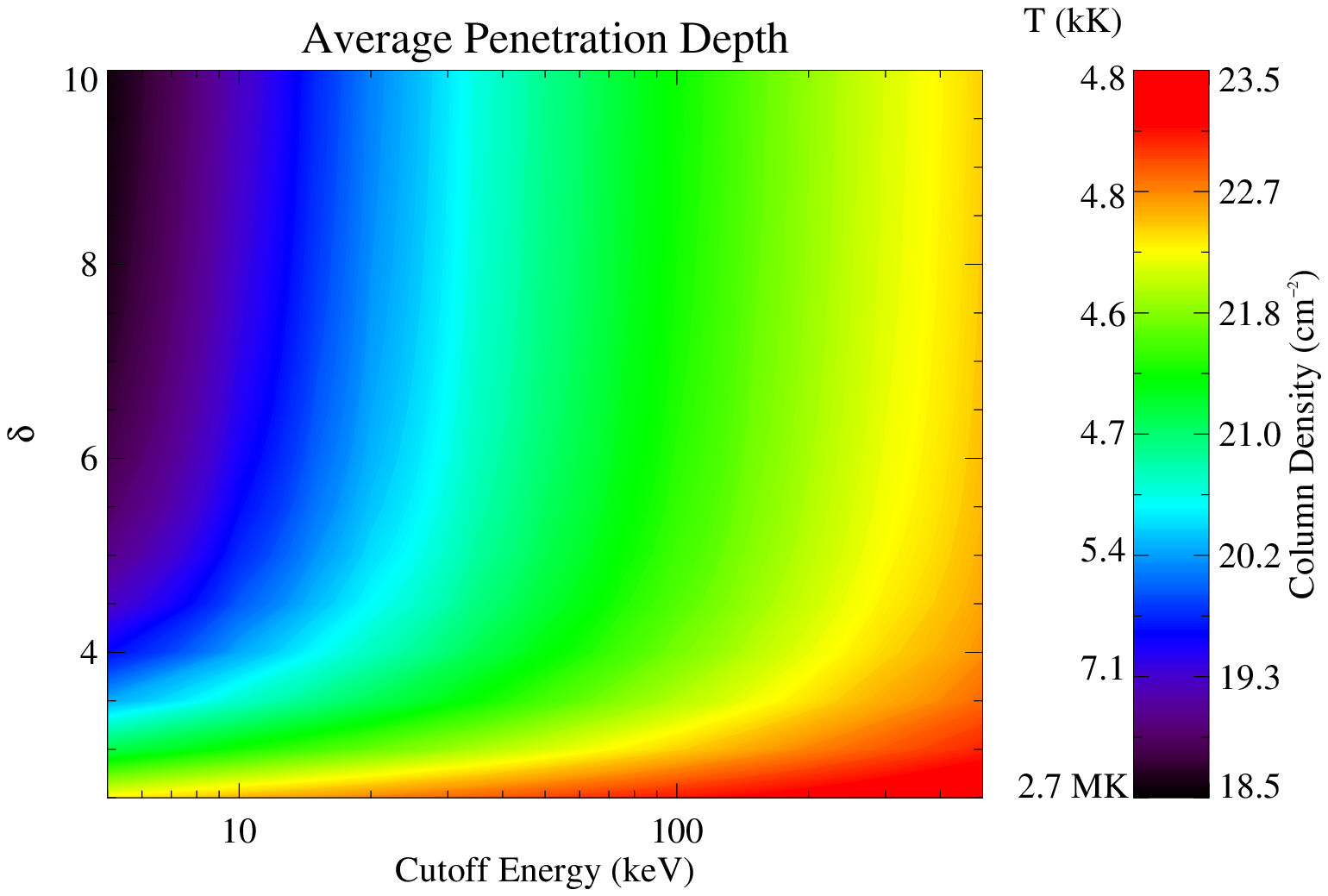}{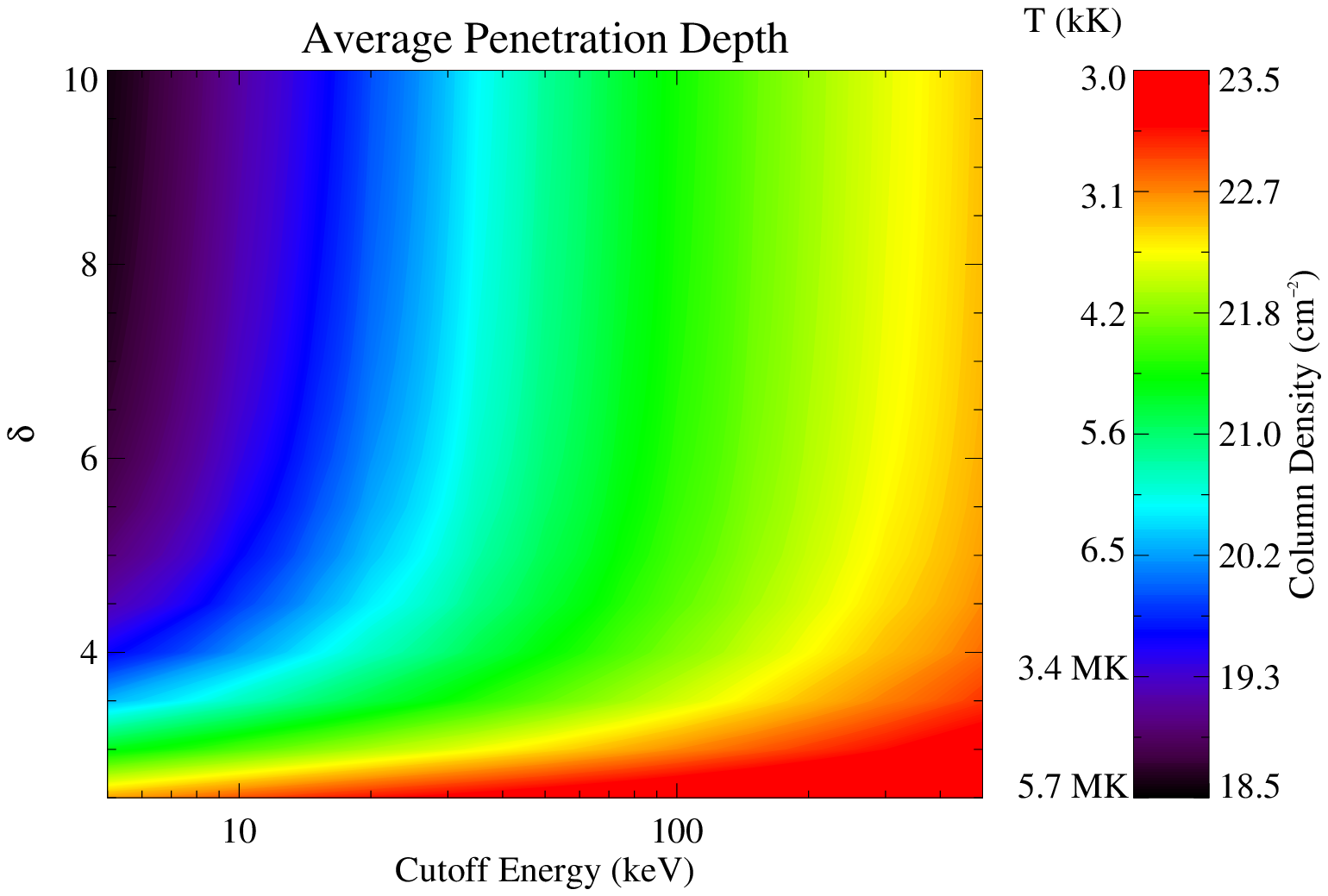}
 \caption{The penetration depth as a function of cutoff energy and spectral index, $\delta$, in the QS.SL.HT atmosphere (left) and M dwarf atmosphere (right).
 \label{fig:aveh}}
\end{figure*}

First, we consider how varying $E_c$ and $\delta$ affect the location of beam impact. Figure~\ref{fig:avehexamp} shows a few representative examples of the heating rate due to collisions with beam particles as a function of hydrogen column density in the QS.SL.HT loop. Clearly the heating extends over a range of column densities. We have determined an average value indicated by the dashed lines. In the following discussion, the term penetration depth (in units of column density) refers to this average. Figure~\ref{fig:aveh} shows the beam penetration depth as a function of cutoff energy and the power-law index in the QS.SL.HT and M dwarf flux tubes. A striking result of this analysis is how strongly dependent penetration depth is on $\delta$. Low values of $\delta$ imply a significant number of higher energy particles even for spectra with low cutoff energies. For example, an electron distribution with $E_c = 10$ keV and $\delta = 3.0$ penetrates as deeply as one with $E_c = 70$ keV and $\delta = 5.0$. Interestingly, for distributions that have both very low values of $E_c$ ($< 8$ keV) and high $\delta$ ($> 6$) much of the energy is deposited directly in the corona. 

\begin{figure}
 \ifeapj \hspace*{-.5in} \epsscale{1.32} \else \epsscale{1} \fi
 \plotone{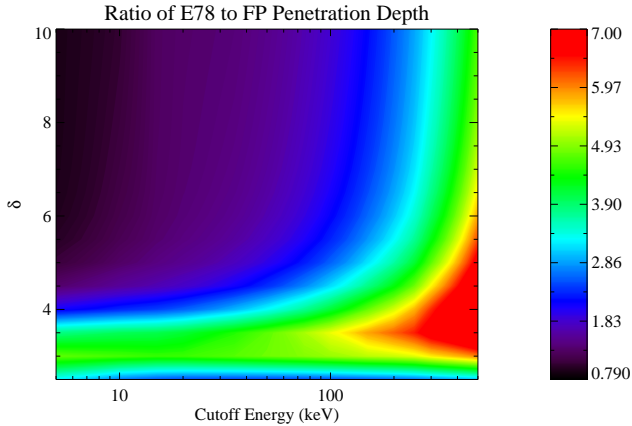}
 \caption{Ratio of the beam penetration depth calculated using our Fokker-Planck technique to that of E78 in the QS.SL.HT loop.
 \label{fig:fpe78}}
\end{figure}

Many previous studies \citep[e.g.,][]{1999ApJ...521..906A, 2005ApJ...630..573A, 2006ApJ...644..484A, 2010ApJ...711..185C} implemented the analytic expression for beam penetration depth derived by E78. That expression includes non-uniform ionization \citep{1994ApJ...426..387H} but does not include relativistic effects. Since the E78 expression has been so widely used, it is informative to compare it with the more complete treatment that we described in Section~\ref{sec:beamheat}. In Figure~\ref{fig:fpe78}, we plot the ratio of beam penetration depths obtained from E78 to that derived by solving the Fokker-Planck equation. We find E78 works quite well in the the range $E_c < 40$ keV and $\delta > 4.5$ but becomes increasingly inaccurate for higher $E_c$ and lower $\delta$. In this regime, E78 predicts a penetration depth as much as 7 times greater. As the cutoff energy increases, relativistic effects become significant. Due to the Lorentz length contraction, relativistic electrons experience a higher ambient density than would be expected from classical theory resulting in less penetration. Relativistic effects become less pronounced with increasing $\delta$, since there are fewer high energy electrons in the distribution.

\begin{figure}
 \ifeapj \hspace*{-.5in} \epsscale{1.32} \else \epsscale{1} \fi
 \plotone{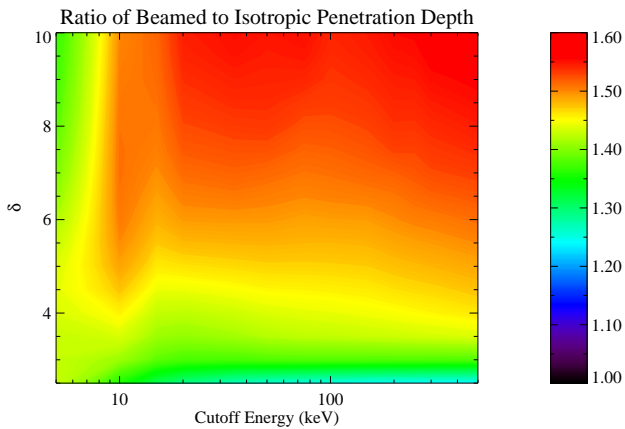}
 \caption{Ratio of the penetration depth for a highly beamed to isotropic injected distribution. Calculated using the QS.SL.HT initial atmosphere. 
 \label{fig:beamiso}}
\end{figure}

Figure~\ref{fig:beamiso} shows the ratio of the penetration depth for highly beamed non-thermal electrons to an isotropic in pitch angle distribution. As expected, the beamed electrons penetrate more deeply. As $\delta$ increases this effect becomes more pronounced. This is because higher energy particles undergo more collisions before stopping, so their initial pitch angle is less important in determining their final stopping depth. A larger $\delta$ means fewer high energy particles are in the distribution, so the initial pitch angle has a more pronounced effect. 

\begin{figure*}
 \ifeapj \epsscale{1.16} \else \epsscale{1} \fi
 \plotone{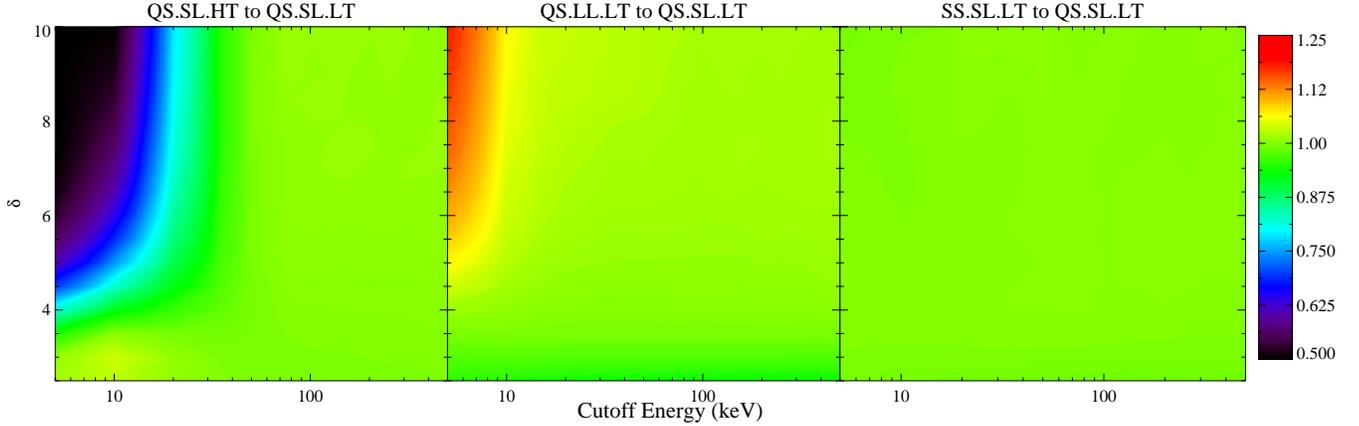}
 \caption{Ratio of the beam penetration depth in the QS.SL.HT (left), QS.LL.LT (middle), and SS.SL.LT (right) atmospheres to the penetration depth in the QS.SL.LT simulation as a function of cutoff energy and spectral index. 
 \label{fig:compare_atmos}}
\end{figure*}

Figure~\ref{fig:compare_atmos} compares the beam penetration depths as a function of cutoff energy and spectral index for several loops. The left panel shows the ratio of the penetration depths for beams propagating in QS.SL.HT to QS.SL.LT. For low values of the cutoff energy, the beam penetrates to a column mass approximately twice as great in the QS.SL.LT loop. In this loop, the coronal temperature and density are lower than in QS.SL.HT. It is the low energy electrons that are most affected by this difference. The middle panel compares beam penetration depths in QS.LL.LT with QS.SL.LT. In this case, beams with low-cutoff energy penetrate less deeply in QS.SL.LT. Since QS.SL.LT is a shorter loop than QS.LL.LT, the coronal density is greater resulting in less penetration for the lowest energy electrons. The right panel compares QS.SL.LT with SS.SL.LT. These loops are nearly identical in their coronal and upper chromospheric regions. They differ significantly in the photosphere, but very few electrons are able to penetrate to that depth so beam penetration depths are similar in these loops in the range of cutoff energies and spectral indices studied here. However, since they have very different photospheric temperatures, these loops predict significantly different levels of white light emission. This difference will be important in studying the white light emission predicted from flaring loops. 

\subsection{Collision Rates}
\begin{figure}
\ifeapj \epsscale{1.16} \else \epsscale{1} \fi
\plotone{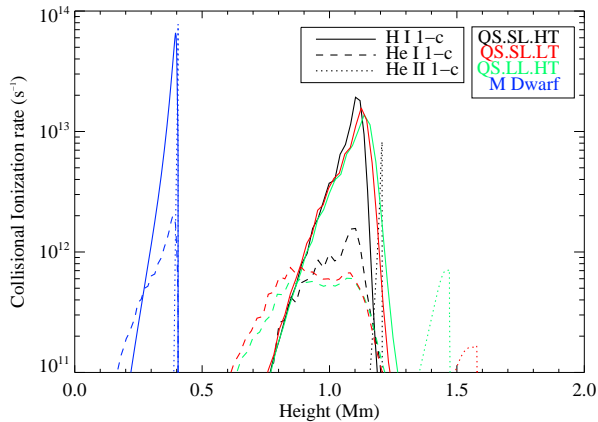}
\caption{The beam induced collisional rates, $n_i C_B^{nt}$, for \ion{H}{1}~1-c (solid line), \ion{He}{1}~1-c (dashed line), and \ion{He}{2}~1-c (dotted line) in the QS.SL.HT (black), QS.SL.LT (red), QS.LL.HT (green), and M dwarf (blue) loops.  In each case the injected electron beam is for $E_c = 20$~keV and $\delta  = 5$. The \ion{He}{2}~1-c rates have been scaled by a factor of 1000 so that they will fit in the plot.\label{fig:compcollrates}}
\end{figure}

During flares, direct excitations and ionizations by beam particles can be much larger than thermal rates in the region of beam impact\footnote{Even though these excitations are important, throughout the remainder of this section, we will focus exclusively on the ionizations. This is because the excitation rates can be obtained simply from scaling the ionization rates by the $R^{H,nt}$ coefficients as seen in Equation~\ref{eqn:hrates}.}. To understand the relative importance of collisional ionizations by flare-accelerated particles, it is informative to compare them with the thermal rates. In Figure~\ref{fig:compcollrates}, we plot the ionization rates, $n_i C_{beam}^{nt}$, in several loops. Here $n_i$ is the ground state number density. We find the ionization rates increase dramatically around beam impact. For example, in the solar loops the \ion{H}{1}~1-c rates peak at $\sn{2}{13}$~s$^{-1}$. To put this into perspective, the thermal rates in this region are $\sim 10^3$ s$^{-1}$ so some 10 orders of magnitude smaller. These results are relatively independent of the initial loop conditions, with the exception of the M dwarf loop. $n_i C_{beam}^{nt}$ increases deeper in the M~dwarf loop compared to the solar loops because in the former the chromosphere forms deeper. 

\begin{figure}
  \ifeapj \hspace*{-.3in} \includegraphics[scale=.52]{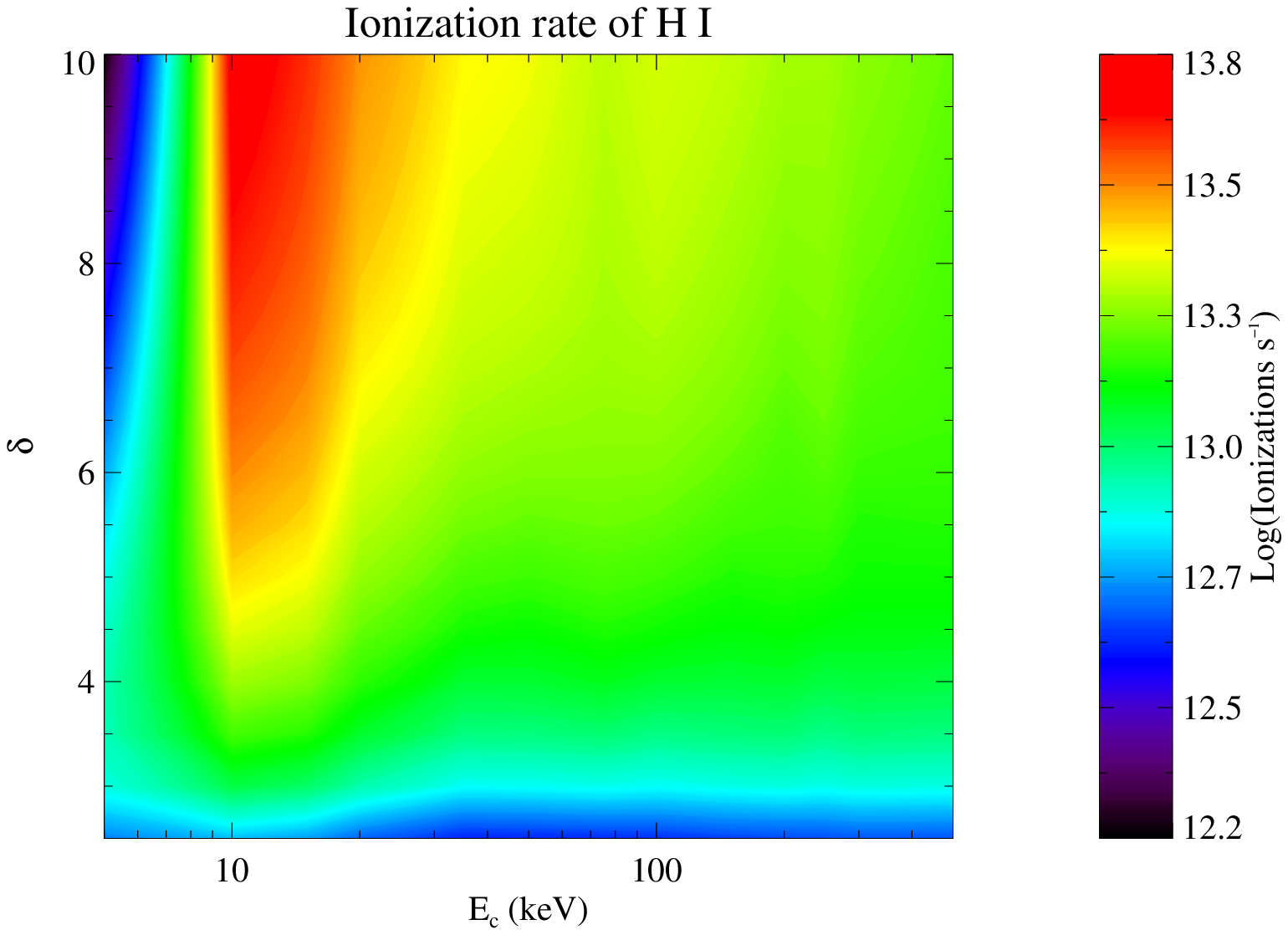} \else \includegraphics[scale=.5]{ch.eps} \newline \fi
  \ifeapj \hspace*{-.3in} \includegraphics[scale=.52]{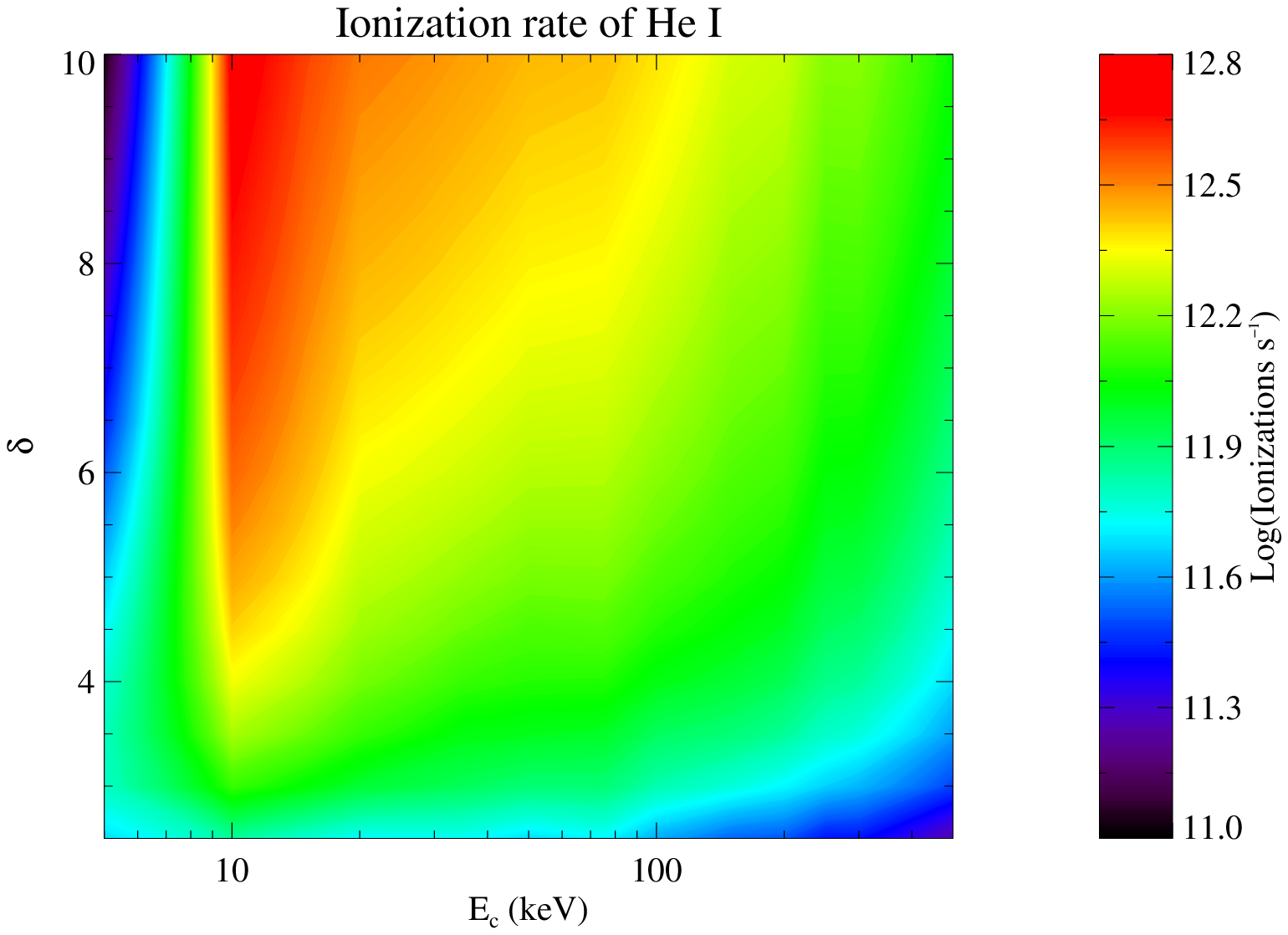} \else \includegraphics[scale=.5]{chei.eps} \newline \fi
  \ifeapj \hspace*{-.3in} \includegraphics[scale=.52]{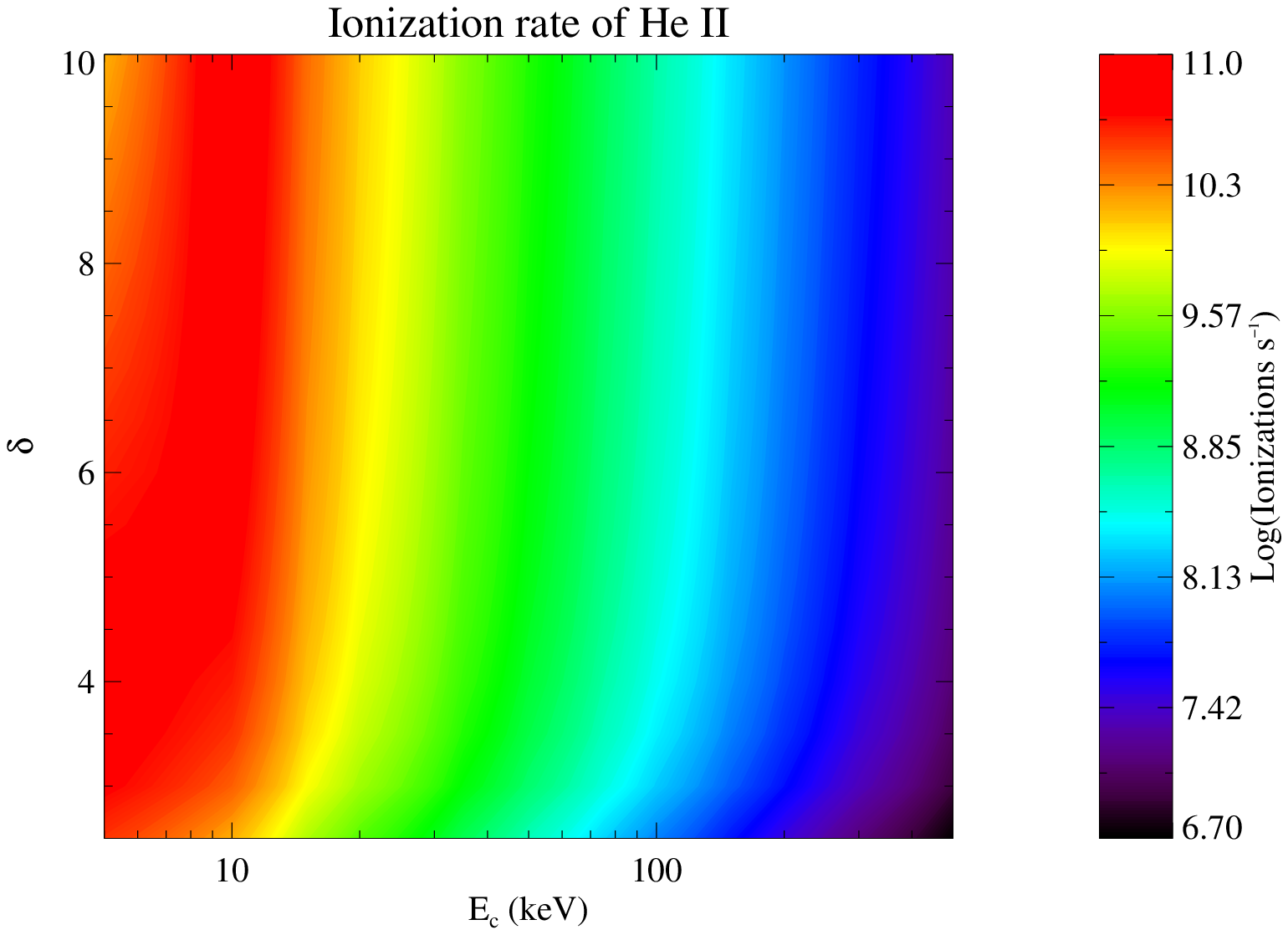} \else \includegraphics[scale=.5]{cheii.eps} \fi
 \caption{Collisional ionziation rates for \ion{H}{1} (top), \ion{He}{1} (middle), and \ion{He}{2} (bottom) produced by non-thermal particles injected at the top of the QS.SL.HT loop. \label{fig:collrates}}   
\end{figure}

It is useful to model how these collision rates vary with injected electron beam spectra. Figure~\ref{fig:collrates} shows the peak non-thermal collisional ionziation rates for electron beams injected into the QS.SL.HT loop as a function of $E_c$ and $\delta$. \ion{H}{1}~1-c and \ion{He}{1}~1-c rates have similar dependence on these parameters, peaking at lower $E_c$ and higher $\delta$. $Q_{beam}$ is narrower and higher peaked in this regime (see the middle and bottom panels of Figure~\ref{fig:avehexamp}) resulting in a higher peak collision rate. However, the collision rates drop quickly for very low $E_c$. There much of the beam is stopped in the transition region and corona where $n_i$ is low. The \ion{He}{2}~1-c rates rapidly decrease for increasing $E_c$. A beam with larger $E_c$ penetrates to a deeper, cooler, and hence lower \ion{He}{2} density region of the atmosphere. 

\subsection{Return Current Heating in the Corona}
\begin{figure}
 \ifeapj \hspace*{-.2in} \epsscale{1.15} \else \epsscale{1} \fi
 \plotone{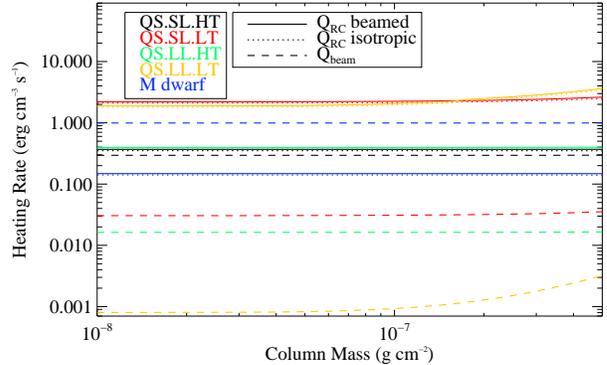}
 \caption{Volumetric heating rates for the return current (solid and dotted lines) and electron beam (dashed lines) in the QS.SL.HT (black), QS.SL.LT (red), QS.LL.HT (green), QS.LL.LT (orange), and M dwarf loops (blue). The solid and dashed lines indicate return current heating for beamed and isotropic distributions of electrons, respectively. These rates are all calculated assuming $E_c = 20$ keV and $\delta = 5$. \label{fig:rcbh}}
\end{figure}

To understand how return currents are likely to affect flare dynamics, it useful to compare their heating rates ($Q_{rc}$) to the heating rates produced directly by collisions ($Q_{beam}$). In the following discussion, it should be kept in mind that we are comparing only \emph{coronal} heating rates -- not those in the chromosphere where $Q_{beam}$ dominates. A comparison of these rates in several loops is shown in Figure~\ref{fig:rcbh}. In the QS.SL.HT loop, $Q_{rc}$ and $Q_{beam}$ are similar in size. However, in all other cases, $Q_{rc}$ dominates by at least an order of magnitude, being stronger in the cooler loops. This is because the resistivity is strongly and inversely dependent on the coronal temperature. Hot loops have much less resistance so are heated by $Q_{rc}$ less. Therefore, we speculate that return currents are likely to be most important in the early phases of flares before the corona has been heated to very high temperatures. This figure also compares $Q_{rc}$ for the case of highly beamed electrons (solid lines) and isotropically injected electrons (dotted lines). These rates are similar in all cases indicating that the pitch-angle distribution of the injected electrons makes little difference in $Q_{rc}$.

\begin{figure}
 \ifeapj \hspace*{-.3in} \epsscale{1.25} \else \epsscale{1} \fi
 \plotone{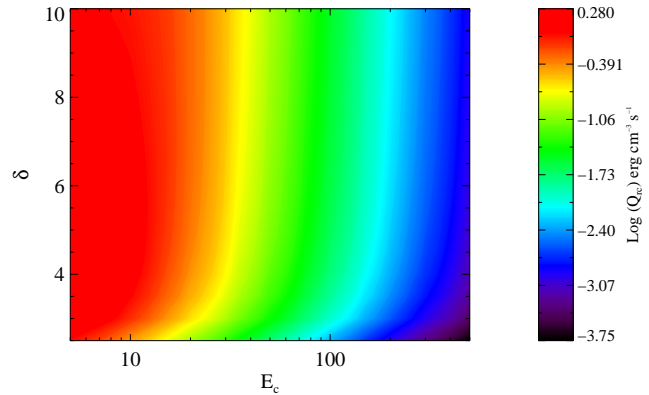}
 \caption{Volumetric return current heating rate as a function of $E_c$ and $\delta$ in the QS.SL.HT loop. \label{fig:rc}}
\end{figure}

From the results of this parameter study, we can also compare how $Q_{rc}$ varies with $E_c$ and $\delta$. Figure~\ref{fig:rc} shows the volumeteric heating rate due to return current in the corona of the QS.SL.HT loop as a function of these parameters. The same energy flux ($10^{10}$ erg cm$^{-2}$ s$^{-1}$) was injected for every simulation in this study, so simulations with higher $E_c$ have fewer total injected electrons. Return current heating scales with the particle flux ($F_e$) so fewer particles results in less heating. The energy per particle factors into $F_e$ only through its velocity, so the return current heating rate is only weakly dependent on $\delta$. 

\section{Conclusions}\label{sec:conc}
We have developed a computational framework capable of modeling the evolution of magnetic flux loops in response to flare heating. This model can be applied to loops on the Sun and on dMe flare stars, and it includes many of the physical processes most important to flare dynamics. Our code solves the optically-thick, non-LTE radiative transfer equation for many transitions important in the chromosphere where much of the flare energy is deposited allowing us to model flare emission which can be compared directly to observations. We have implemented a method for solving the Fokker-Planck equation describing how flare-accelerated particles interact with the ambient plasma and have used that to model the heating, momentum deposition, collisional excitations and ionizations, and return currents produced by these particles.

With our model, we have performed a parameter study to understand how these phenomena depend on the energy spectrum of a beam of flare-accelerated particles. We have assumed a power-law energy distribution for these particles and found that their penetration depth is strongly dependent on the beam cutoff energy and power-law index and more weakly dependent on their initial pitch-angle distribution. For distributions with low power-law index or high cutoff energy, relativistic effects become important causing a penetration to depths as much as $7$ times less than predicted by classical theory. Varying the conditions of the loops (i.e., the coronal temperature and density and photospheric temperature) has a small effect on the penetration depth for distributions with low cutoff energy. The collisional ionizations produced by impacts from these flare-accelerated particles also are strongly dependent on cutoff energy and power-law index. They can be 10 orders of magnitude greater than thermal collision rates. The beam particles also induce a return current which deposits heat in the corona. We have found that, depending on coronal conditions, return current heating in the corona can be more than an order of magnitude greater than coronal beam heating. 

In this parameter study, we have focused on understanding the effects of flare-accelerated electrons. In future work, we will extend this to flare-accelerated ions which are also known to be important to flare energetics. This work has focused on the initial response of flaring loops. However, a key feature of our framework is its ability to model the dynamical evolution of these loops. In future work, we will present a parameter study exploring their evolution in response to flare-accelerated particles. 
\acknowledgments
This work has been supported by grants through the NASA Heliophysics Supporting Research and Technology and the NASA Living With a Star programs. This research was aided by many useful discussions with the participants of the Chromospheric Flares meeting held at the International Space Science Institute (ISSI) in Bern, Switzerland. AFK acknowledges the support of the NASA Postdoctoral Program at the Goddard Space Flight Center, administered by Oak Ridge Associated Universities through a contract with NASA, and from support through UMCP GPHI Task 132. The research leading to these results has received funding from the European Research Council under the European Union's Seventh Framework Programme (FP7/2007-2013) / ERC grant agreement nr 291058 and by the Research Council of Norway through the grant "Solar Atmospheric Modelling" and through grants of computing time from the Programme for Supercomputing. We thank G. Holman for helpful discussions about return current heating. 
\FloatBarrier
\bibliography{radynpaper.bib}

\end{document}